%Paper: hep-th/9306030
%From: fiore@tsmi19.sissa.it
%Date: Fri, 04 Jun 1993 19:05:58 +0100

%corrected/revised version of the paper
\magnification=1200

\rm \font \myfont=cmr10
\font\large=cmr10 at 14.4truept
\font\mysmall=cmr9

\myfont\parindent=0pt
\hfill{June 1992}
\baselineskip=15pt
\vskip 1cm
%\centerline
\vskip1.3truecm
\centerline{ Gaetano Fiore }
\vskip0.3truecm
\centerline{SISSA, Strada Costiera 11, 34014 Trieste,Italy}
\vskip0.3truecm
\centerline{and}
\vskip0.3truecm
\centerline{Istituto Nazionale di Fisica Nucleare, Sezione di Trieste}
\vskip1.0truecm
{\bf THE $SO_q(N,{\bf R})$-SYMMETRIC HARMONIC OSCILLATOR ON THE QUANTUM}
\centerline{\bf EUCLIDEAN SPACE ${\bf R}_q^N$ AND ITS HILBERT SPACE STRUCTURE}
\vskip1.3truecm

\centerline{ABSTRACT}

\midinsert
\font\mysmall=cmr9
{\mysmall
We show that the isotropic harmonic oscillator in the ordinary euclidean
space ${\bf R}^N$ ($N\ge 3$) admits a natural
q-deformation into a new quantum mechanical
model having a q-deformed symmetry (in the sense of quantum groups),
$SO_q(N,{\bf R})$. The q-deformation is the consequence of replacing ${\bf
R}^N$
by ${\bf R}^N_q$ (the corresponding quantum space). This
provides an example of quantum mechanics on a noncommutative geometrical space.
To reach the goal, we also have to deal with a sensible definition of
integration over ${\bf R}^N_q$, which we use for the definition of the scalar
product of states.

}
\endinsert
\vskip1.3truecm

\vfill\eject

\centerline{\bf {\large 1.} Introduction}

{}~~

{}~~~The development of Physics has been often characterized by the
introduction
of some more general and accurate theory as sort of deformations of already
known and accepted ones. A well-known example is special relativity, which can
be viewed as a deformation of Galileo's relativity; the velocity of light
plays the role of defomation parameter. Another example is quantum mechanics,
which can be seen as a deformation of classical mechanics, Planck constant
being the deformation parameter.

{}~~~ Within the recent increasing interest for quantum
groups the question has been raised [1] whether these fascinating
mathematical objects can replace (or generalize) Lie groups in the description
of the fundamental symmetries of physics, since they can be considered as
continuous deformations of Lie groups themselves [2],[3].
One may ask whether the axioms of quantum mechanics are compatible with a
more general description of continuous symmetries than the usual one,
i.e. the one provided by the theory of Lie groups and of their unitary
representations over the Hilbert
spaces of physical states; many new possibilities in this direction seem
to be open [4].

{}~~~In particular it looks tempting to consider deformations of the symmetries
of space(time) [5],[6],[7],[8],[9]; in such a case quantum groups
and/or the underlying quantum spaces [3] replace classical space(time) and
represent examples of noncommutative geometries [10]. Such
geometries look promising for describing the microscopic structure
of spacetime.

{}~~~In this work we consider a specific finite-dimensional quantum mechanical
model with a symmetry Lie group $G$ and ask whether it admits a q-deformation
($q\equiv$parameter of deformation) such that the symmetry of the q-deformed
model be described by the corresponding  quantum group $G_q$. Actually, in the
deformation not only we should replace the Lie group symmetry of the system
(i.e. of its hamiltonian) by the quantum group symmetry, but first we should
replace the concept of covariance of physical laws w.r.t. a Lie group by the
concept of their covariance w.r.t. a quantum group.

{}~~~In fact we are going to build the (time independent)
harmonic oscillator on the $N$-dimensional real quantum euclidean
space (we will call it ${\bf R}^N_q$, and $N\ge 3$) as the deformation of the
classical isotropic
harmonic oscillator on ${\bf R}^N$. Correspondingly, the symmetry group
$SO(N,{\bf R})$ of rotations is deformed into the quantum group
$SO_q(N,{\bf R})$. The construction of the q-deformed quantum model
is performed in `` coordinate representation ''.

{}~~~As known, there are two dually related ways to look at Lie groups, at the
spaces of their representation, and to q-deform them (see for instance [3]).
In the first case one considers a Lie group $G$ (or its Lie algebra ${\cal G}$)
and its action on a representation space $V$; in this case one deforms the
universal enveloping algebra ${\cal U}({\cal G})$ of ${\cal G}$
and its representations. In the second case, one
considers the Hopf algebra $Fun(G)$ of functions on the group $G$ and its
corepresentations; then deformation involves the latter objects.
The present work is based on the second approach.

{}~~~It is convenient to recall how one can formulate covariance of the
physical
description using the language of coaction and corepresentations; we do this
job in the specific case in which we consider a point particle in ordinary
3-dimensional space ${\bf R}^3$ and we take the group of rotations
$SO(3,{\bf R})$ of ${\bf R}^3$ as the symmetry group $G$.

{}~~~$Fun(G)$ is the commutative algebra of functions
on the group $G:=SO(3,{\bf R})$. The functions can be expressed as power series
in the basic variables $T^i_j\in Fun(G)$,
$i,j=1,2,3$, which are defined by $T^i_j(g)=g^i_j$ ($g\in G$ and
$\Vert g^i_j\Vert \in adj(G)$,
where $adj(G)$ denotes the adjoint representation of $G$). Any vector
$\bar V\equiv(V^i)$ provides a fundamental corepresentation of
the (left) coaction $\phi_L$ of $Fun(G)$:
$$
\phi_L(V^i):=T^i_j\bigotimes V^j                                 \eqno (1.1)
$$
For instance if $\bar V=\bar X$ ($\bar X\equiv$the position operator of the
quantum particle in a reference frame $S$), then the position operator
$\bar X'$ of the particle in the frame $S'$ obtained from $S$ by a rotation
$g$ will be given by
$$
[\phi_L(X^i)](g,\cdot):=[T^i_j\bigotimes X^j](g,\cdot)=
T^i_j(g)X^i=g^i_jX^j=X^{'i}.                                  \eqno (1.2)
$$
$\phi_L$ is extended as an algebra homomorphism to higher rank
corepresentations of $Fun(G)$. For instance:
$$
\phi_L(X^iX^j):=\phi_L(X^i)\phi_L(X^j).                        \eqno (1.3)
$$
This implies for the commutator $[X^i,X^j]$
$$
\phi_L([X^i,X^j])=T^i_hT^j_k\bigotimes X^hX^k-
T^j_kT^i_h\bigotimes X^kX^h.                                \eqno (1.4)
$$
This formula is consistent with the commutation relations
$$
[X^i,X^j]=0,~~~~~~~~~~~~~~[T^i_h,T^j_k]=0                   \eqno (1.5)
$$

{}~~~It is easy to see from formula (1.4) that deformations of the commutation
relations $(1.5)_a$ and $(1.5)_b$ are strictly coupled; the use of the quantum
space ${\bf R}^3_q$ and of the quantum group $SO_q(N,{\bf R})$ lets one perform
a consistent q-deformation of both.

{}~~~To understand the line of development of the present work, let us
briefly review the basic mathematical tools which allow the formulation of
classical (i.e. undeformed) quantum mechanics in the coordinate representation
$\Pi$ over the $N$-dimensional space ${\bf R}^N$. We can summarize its main
ingredients in the following list (with self-evident notation):

{}~~~$\bullet~1)$ There exists a vector calculus on ${\bf R}^N$ which is
covariant w.r.t. the group $SO(N,{\bf R})$ of rotations of ${\bf R}^N$.
$x\equiv (x^i)\in {\bf R}^N$ is the coordinate vector.

{}~~~$\bullet~2)$ There exists a (SO(N,{\bf R})-covariant)
differential calculus $D$ on ${\bf R}^N$ ($derivatives\equiv\partial^i \in D$).

{}~~~$\bullet~3)$ There exists an antilinear involutive antihomomorphism
defined
on the algebra of functions of $x,\partial$, the socalled complex
conjugation $*$.

{}~~~$\bullet~4)$ The vectors belonging to the Hilbert space ${\cal H}$ are
represented by
$$
\Pi:|u>\in {\cal H}\rightarrow\psi_u(x)\in {\cal L}^2({\bf R}^N)
$$

{}~~~$\bullet 5)$ Relevant operators (observables etc.) are represented in
terms
of functions of $x\cdot,\partial$. Eigenvalue equations
are represented by differential equations (at least in a domain dense in
${\cal L}^2({\bf R}^N)$.

{}~~~$\bullet 6)$ scalar products are evaluated by means of Riemann
integration,
$$
<u|v>=\int d^Nx~\psi_u^*\psi_v,
$$
which satisfies Stoke's theorem and therefore automatically makes the momentum
operators ${1\over i}\partial^i$ hermitean.

{}~~~$\bullet 7)$ The Schroedinger equation for the harmonic oscillator on
${\bf R}^N$ admits an algebraic solution by means of the creation and
destruction operators (which are also represented using $x\cdot,\partial$)

{}~~~To construct the q-deformed model we use a q-deformed version of each of
these ingredients.
The analogs of points 1), 2), 3) are thoroughly developed in Ref. [3],[11],
[12],[13];
the analogs of the remaining points are essentially new. They are
constructed here using
some partial results given in Ref. [14]. Using these q-deformed
tools, we show that a sensible q-deformed harmonic oscillator on ${\bf R}^N_q$
(with symmetry $SO_q(N,{\bf R})$) can be constructed. In other words we
will show that such a model satisfies the fundamental
axioms of quantum mechanics.

{}~~~The plan of the work is as follows.

{}~~~Section 2. is a summary (based essentially on Ref. [3],[11],[14]) of
already presented results concerning the
quantum group $SO_q(N,{\bf R})$, the quantum space ${\bf R}^N_q$,
the (two) $SO_q(N,{\bf R})$-covariant differential calculi $D,\bar D$ on
${\bf R}^N_q$, the (time-independent) Schroedinger equation of the harmonic
oscillator on ${\bf R}_q^N$ and its algebraic solution. The Schroedinger
equation is formulated here in terms of the q-deformed laplacians of $D,\bar D$
and it is solved using a suitable generalization of the
classical creation/destruction
operators. The spectrum is bounded from below, as physics requires.
The eigenfunctions are the ``q-deformed'' Hermite functions.

{}~~~Sections 3.,4. deal with the definition of integration over ${\bf R}^N_q$.
Integration is thoroughly defined using Stoke's theorem only on some
functions of the type
$polynomial\cdot gaussian$; the latter will be involved in the definition of
the
scalar products of states of the harmonic oscillator.
To define integration in full generality one should carefully
delimit the domain of functions on which commutation of integration and
infinite
sums makes sense; this is out of the scope of this work.
In Sect. 3 we analyse the desired requirements that an honest definition of
integration should satisfy; among them Stoke's theorem plays a special
role. In Sect. 4 and appendix B we carry
out the construction of the integral for the abovementioned relevant
fucntions; at the end of that section we comment on a surprising feature
regarding the behaviour of integration under dilatation of the
integration variables, a sort of ``quantized'' scaling invariance.

{}~~~In the remaining sections we construct the Hilbert space of the harmonic
oscillator.
First a pre-Hilbert space ${\cal H}$ is introduced by representing the states
in two different ways inside the space $Fun({\bf R}^N_q)$ (Sect. 5), so as to
construct in the simplest possible way hermitean operators as functions of
coordinates and derivatives. The
two representations $(\Pi,\bar \Pi)$ correspond respectively to $D,\bar D$.
It is shown that the position/momentum operators and the
hamiltonian of the harmonic oscillator are observables, i.e. hermitean
operators. In Sect. 6 we find some observables commuting with the hamiltonian,
namely the
angular momentum components and the square angular momentum, and we determine
the spectrum and eigenfuntions of the latter. With the help
of these results we prove (Sect. 7) the positivity of the scalar product
introduced in Sect. 5. This allows the completion of ${\cal H}$ into a Hilbert
space $[{\cal H}]$. Section 8 contains the conclusions of the present work.

{}~~~In most cases any relation admitting a barred and an unbarred version will
be
explicitly written only in the unbarred representation, the barred version
being immediately available after some simple replacements (see Section 2.).

{}~~~q-deformed harmonic oscillators have already been treated by other authors
[15] starting from a purely algebraic approach, in the sense that
creation/destruction operators
with some prescribed commutations relation are postulated from the very
beginning without any reference to a geometrical framework. The deformation
considered there concerned the well-known hidden $su(n)$ symmetries of the
harmonic oscillator hamitonians. Here and in [16], on the contrary,
the deformation concerns the rotation symmetry of the space itself: in other
words, a geometrical framework is the
starting point and creation/destruction operators are constructed out of the
deformed `` coordinates '' and `` derivatives ''.

{}~~

{}~~

\centerline {\bf {\large  2.} Notation and Preliminary results}

{}~

{}~~~In Ref. [11] the differential calculus on the
$N$-dimensional ($N \ge 3$) real quantum
euclidean space (we will call it ${\bf R}_q^N$ in the sequel)
was developed along the successful line already
followed in Ref. [7], the
guiding principles being essentially Leibniz rule together with nilpotency
for the exterior derivative, and the requirement of covariance with respect
to the quantum group $SO_q(N,{\bf R})$ for the whole calculus.
As noticed by its authors, the differential calculus
explicitly developed there was one of the two possible (linearly independent)
versions; the second one can be obtained from the first by very
simple replacements. Under complex conjugation each of the two calculi is
mapped
into the other. In a recent work [13] it is shown that these two calculi
generate the same ring, which we will call $Diff({\bf R_q^N})$ in the sequel,
namely barred derivatives can be expressed as some nonlinear (and quite
complicated) functions of
coordinates and unbarred ones, and so on.
For the line of development of this work we don't need this result, and we
will essentially treat the two calculi as independent objects.

{}~~~In this section we first recollect some
basic definitions and relations characterizing the quantum group
$SO_q(N,{\bf R})$, the algebra $Fun({\bf R_q^N})$ of
functions on ${\bf R}_q^N$ (which is generated by the noncommuting
coordinates $x~(=\{x^i\},~i=1,...,N)$), its two differential calculi (which
will
be denoted by $D$,$\bar D$) and the action of the complex conjugation $*$
on all of them. For further details we refer the reader to [3], [11],[13].
Then we summarize the results of [14] concerning the Schroedinger equation
of the harmonic oscillator on ${\bf R}^N_q$ and its solution.

{}~

{\bf 2.1.  The Real Quantum Euclidean Space ${\bf R}^N_q$
and its Two Differential Calculi}

{}~

{}~~~The real section $SO_q(N,{\bf R})$ ($q\in {\bf R}^{+})$ of the quantum
group
$SO_q(N)$ [3] is taken as the symmetry quantum group of the whole construction.
As known, elements of the Hopf algebra $Fun(SO_q(N))$ (the algebra of
`` functions '' on the quantum group $SO_q(N)$) are formal ordered power series
in the generating elements $\{T^i_j\},~i,j=1,2,...,N$. The latter satisfy the
relations
$$
TCT^t={\bf 1}_{SO_q(N)}C=T^tCT                                   \eqno (2.1)
$$
$$
\hat R(T\otimes T)=(T\otimes T) \hat R.                          \eqno (2.2)
$$
Here ${\bf 1}_{SO_q(N)}$ is the unit element of $Fun(SO_q(N))$,
$C=\Vert C_{ij}\Vert$ is the metric matrix (which is its own inverse,
$C^{-1}=C$), and $\hat R=\Vert\hat
R^{ij}_{hk}\Vert$ is the braid matrix, defined on ${\bf C}^N\otimes
{\bf C}^N$; $\hat R$ is symmetric: $\hat R^T=\hat R$. Both $C$ and $\hat R$
depend on $q$ and are real for
$q\in {\bf R}$. $\hat R$ satisfies the Yang-Baxter equation (in the " braid "
version)
$$
\hat R_{12} \hat R_{23} \hat R_{12} = \hat R_{23} \hat R_{12} \hat R_{23},
                                                             \eqno (2.3)
$$
and admits the very useful decomposition
$$
\hat R_q = q {\cal P}_S - q^{-1} {\cal P}_A +q^{1-N}{\cal P}_1~~~~~~
\hat R_q^{-1} = q^{-1}{\cal P}_S - q {\cal P}_A + q^{N-1}{\cal P}_1.
                                                             \eqno (2.4)
$$
${\cal P}_S,{\cal P}_A,{\cal P}_1$ are the projection operators
onto the three eigenspaces of $\hat R$ (the latter have respectively dimensions
${{N(N+1)}\over 2}-1,{{N(N-1)}\over 2},1$): they project the tensor
product $x\otimes x$ of the fundamental corepresentation $x$ of $SO_q(N)$
into the corresponding irreducible corepresentations (the symmetric,
antisymmetric and singlet, namely the q-deformed
versions of the corresponding ones of $SO(N)$). The projector ${\cal P}_1$ is
related to the metric matrix $C$ by ${\cal P}_{1~hk}^{~~ij}={C^{ij}C_{hk}\over
Q_N}$ (the factor $Q_N$ is defined by $Q_N:=C^{ij}C_{ij}$). $\hat R^{\pm 1},C$
satisfy the relations
$$
C_{mi}\hat R^{\pm 1~ij}_{~~~hk}=\hat R^{\mp 1~jn}_{~~~mh}C_{nk}   \eqno (2.5)
$$
As direct consequences of (2.2),(2.3),(2.5), for any polynomial
$f(t)\in {\bf C}(t)$ we find
$$
f(\hat R)(T\otimes T)=(T\otimes T) f(\hat R),                     \eqno (2.6)
$$
$$
f(\hat R_{12})\hat R_{23}\hat R_{12}=\hat R_{23}\hat R_{12}f(\hat R_{23}),
                                                              \eqno (2.7)
$$
$$
[f(\hat R), P\cdot (C\otimes C)]=0                            \eqno (2.8)
$$
($P$ is the permutator: $P^{ij}_{hk}:=\delta^i_k\delta^j_h$);
in particular this holds for $f(\hat R)=\hat R^{\pm 1},{\cal P}_A,{\cal P}_S,
{\cal P}_1$.

{}~~~ The algebra $O_q^N$ (in the notation of Ref. [3])
is defined as the space of formal series in the ordered
powers of the $\{x^i\}$ variables, modulo the relations
$$
{\cal P}_{A~hk}^{~~ij}x^h x^k =0.                              \eqno (2.9)
$$
For instance, for $N=3$ eq.'s (2.9) amount to the three independent relations
$$
x^1x^2-qx^2x^1=0,~~~~~x^2x^3-qx^3x^2=0,~~~~x^1x^3-x^3x^1+(q^{1\over 2}-q^{-1
\over 2})(x^2)^2=0.                                        \eqno (2.10)
$$
 For q=1 and any $N$ ${\cal P}_{A~hk}^{~~ij}={1\over 2}(\delta^i_h\delta^j_k-
\delta^i_k\delta^j_h)$ so that the $x^i$  coordinates
become commuting variables and their order
in each monomial doesn't matter any more (classical geometry).

{}~~~The exterior derivative $d$ of the differential calculus $D$
 maps the space $\Lambda_q^p(O^N)$ of p-forms into
the one $\Lambda_q^{p+1}(O^N)$ of (p+1)-forms, is nilpotent and
satisfies Leibniz rule:
$$
d^2=0,~~~~~~~~~~~~~~~d\alpha_p| := (d\alpha_p - (-1)^p \alpha_p d ) \in
\Lambda_q^{p+1}, ~~~~~~~~~\alpha_p \in \Lambda_q^p.                 \eqno
(2.11)
$$
In particular if $f\in O^N_q=\Lambda_q^0$, $df|$ is a 1-form. We denote by
$\xi^i := dx^i|$ the exterior
derivatives of the basic coordinates $x^i$; $\{ \xi^i\}$
is a basis of $\Lambda_q^1$. The decomposition $d=\xi^i \partial_i$ defines
the derivatives $\partial_i$
corresponding to each coordinate $x^i$. Indices are raised
and lowered through the metric matrix $C$,  for instance
$$
\partial_i=C_{ij}\partial^j,~~~~~~\partial^i=C^{ij}\partial_j,~~~~C^{ij}:=
(C^{-1})_{ij}=C_{ij}                         \eqno (2.12)
$$
Among the " commutation " relations between $x^i$'s, $\xi^i$'s, $\partial^i$'s
we mention the following:
$$
x^i\xi^j=q\hat R^{ij}_{hk}\xi^hx^k         \eqno (2.13)
$$
$$
\partial^i x^j = C^{ij}+q\hat R^{-1~ij}_{~~hk} x^h\partial^k \eqno   (2.14)
$$
$$
{\cal P}_{A~hk}^{~~ij}\partial^h \partial^k = 0         \eqno (2.15)
$$
{}~~~Higher degree forms can be defined as wedge products
of 1-forms: the wedge products of the basic  1-forms $\xi^i$'s are defined
as their tensor products modulo the relations
$$
{\cal P}_S(\xi \otimes \xi)=0~~~~~~~{\cal P}_1(\xi \otimes \xi)=0. \eqno(2.16)
$$
The wedge product is denoted by $\wedge$ as usual. Therefore
 ${\cal P}_S(\xi \wedge \xi)=0,~{\cal P}_1(\xi \wedge \xi)=0$.

{}~~~The (left) coaction $\phi_L: O_q^N\rightarrow Fun(SO_q(N))\bigotimes
O_q^N$
of $SO_q(N)$ is defined on the basic variables $x^i, \partial^i,\xi^i$ by
$$
\phi_L\circ d=({\bf 1}_{SO_q(N)}\bigotimes d)\circ \phi_L
$$
$$
\phi_L:x^i \rightarrow T^i_j \bigotimes x^j,~~~~\phi_L:\partial^i \rightarrow
T^i_j \bigotimes \partial^j,~~~~\phi_L:\xi^i \rightarrow T^i_j \bigotimes
\xi^j                                                        \eqno (2.17)
$$
and is extended as an homomorphism. The  square length  $xCx:= x^i C_{ij}x^j$
and the laplacian $\Delta := \partial^i \partial_i =
\partial^i C_{ij} \partial^j$ are central elements
respectively in $O_q^N$ and in the algebra of the $\partial$ derivatives; they
are scalars under the coaction of the quantum group:
$$
\phi_L(xCx)= {\bf 1}_{SO_q(N)}\bigotimes (xCx),~~~~~\phi_L(\Delta)=
{\bf 1}_{SO_q(N)}\bigotimes \Delta.                                \eqno (2.18)
$$

{}~~~The above relations define the differential calculus $D=\{d,~\xi^i,~
\partial^i \}$. The barred calculus $\bar D=\{\bar d,~\bar\xi^i,\bar
\partial^i\}$ can be obtained from the unbarred one replacing $d,\xi^i,
\partial^i,\Delta, \Lambda_q,q,\hat R_q$ by $\bar d, \bar \xi^i,
\bar \partial^i, \bar \Delta, \bar \Lambda_q, q^{-1}, \hat R_q^{-1}$.
For instance:
$$
\bar\partial^i x^j = C^{ij}+q^{-1}\hat R^{ij}_{hk} x^h \bar\partial^k
                                                             \eqno (2.19)
$$

{\bf Note:} in the sequel we will usually omit relations and definitions
concerning the barred calculus/representation. Once and for all, we inform
the reader that they can be obtained from the ones concerning the unbarred
calculus/representation through the above replacements and, more
generally, through the replacement ${\cal O}\rightarrow \bar {\cal O}$
for all new objects ${\cal O}$ that we are going to introduce in the work.

{}~

{}~~~In Ref. [13] it is shown that all objects of
$\bar D$ lie in the ring $Diff({\bf R_q^N})$ generated by $D$, namely can be
expressed as some
(nonlinear) functions of the unbarred objects (and viceversa). In the sequel
we will only use the relation
$$
\bar \Delta= q^N G_{q^4}\Delta,                                     \eqno
(2.20)
$$
where $G_q$ denotes the dilatation operator belonging to this ring with action
$$
G_qf(x,\partial, \bar\partial):=f(q^{-1\over 2}x,q^{1\over 2}\partial,q^{1\over
2}\bar\partial),                                                \eqno (2.21)
$$
whose explicit expression in terms of $x,\partial$ is not needed here.

{}~~~If $q\in {\bf R}$ one can introduce an antilinear involutive
antihomomorphism
$*$:
$$
*^2=id~~~~~~~~~~~~~~(AB)^*=B^*A^*                                \eqno (2.22)
$$
on $Fun(SO_q(N)),Diff({\bf R_q^N})$. Since the point $q=0$ is singular for
the $\hat R$
and $C$ matrices, in the sequel we will specialize the discussion to the case
$q\in {\bf R}^+$. On the basic variables $T^i_j$ $*$ is defined by
$$
(T^i_j)^*=C^{li}T^l_mC_{jm},                                    \eqno (2.23)
$$
whereas it maps $x^i,\xi^i,\partial^i,d$ into a combination
of $x^i,\bar \xi^i,
\bar \partial^i, \bar d$ respectively, in the following way
$$
(x^i)^*=x^jC_{ji},~~~~(\xi^i)^*=\bar \xi^j C_{ji},~~~~(\partial^i)^*=
-q^{-N}\bar \partial^j C_{ji},~~~~~d^*=-\bar d.                  \eqno (2.24)
$$
The square lenght turns out to be real, whereas the two laplacians are mapped
one into the other:
$$
(xCx)^*=xCx~~~~~~~~~~~~~~~~~(\Delta)^*=\bar \Delta q^{-2N}.        \eqno (2.25)
$$

{}~~

{\bf Note:} The algebra $Fun(SO_q(N)$ (resp. $O_q^N$), defined by
(2.1),(2.2) (resp. (2.9)), $q\in {\bf R}$, endowed with $*$ is denoted by
$Fun(SO_q(N,{\bf R})$ (resp. $Fun({\bf R_q^N})$) and will be called the
`` algebra of functions on the quantum group $SO_q(N,{\bf R})$ ''
(resp. the `` algebra of functions on the quantum space ${\bf R}_q^N$ '').

{}~

{}~~~We list now some formulae and definition which will useful in the sequel.
For any function $f(x) \in O_q^N$, $\partial^i f$ can be expressed in the
form
$$
\partial^i f= \hat f^i + \tilde f^i_j\partial^j~~~~~~~~\hat f^i,\tilde f^i_j
\in Fun({\bf R_q^N})                                             \eqno (2.26)
$$
upon using (2.14) to move step by step the derivatives to the right of each
$x^i$ variable of each term of
the power expansion of $f$, as far as the extreme right.
Similarly to what has been done in formula (2.11), we denote $\hat f^i$ by
$\partial^i f|$:
$$
\partial^i f|:= \partial^i f - \tilde f^i_j\partial^j (= \hat f^i). \eqno
(2.27)
$$
In an analogous way we can define $\bar \partial^i f|$.

{}~~~The q-exponential function is introduced by
$$
exp_q[Z] :=\sum\limits_{n=0}^{\infty}{Z^n \over (n)_q!},~~~~~~~~~~~~
(n)_q := {{q^n-1}\over {q-1}};                                  \eqno(2.28)
$$
for our scopes its usefulness lies essentially in the relation
$$
\partial^i\{exp_{q^2}[{{\alpha(xCx)}\over\mu}]\}=
\alpha x^i exp_{q^2}[{{\alpha(xCx)}\over \mu}]+
exp_{q^2}[{{q^2\alpha(xCx)}\over \mu}]\partial^i                  \eqno (2.29)
$$
which implies $\partial^i exp_{q^2}[{{\alpha(xCx)}\over\mu}]| \propto x^i
exp_{q^2}[{{\alpha(xCx)}\over\mu}]$.
{}From the definition (2.28) it is easy to check the following q-derivative
property for the exponentials
$$
{{exp_q[qZ]~-~exp_q[Z]}\over {q-1}}~
=~Z~exp_q[Z]                                      \eqno (2.30)
$$

{}~~~Finally, from (2.14), (2.19) it is easy to derive:
$$
\Delta x^i = \mu \partial^i + q^2x^i \Delta~~~~~~~\partial^i(xCx)=\mu x^i +
q^2(xCx)\partial^i                                              \eqno (2.31)
$$
where $\mu:=1+q^{2-N}$. From $(2.31)_a$ it is easy to derive:
$$
\Delta(xCx)^h={\mu^3q^{N+2h-2}\over q^2-1}h_{q^2}(xCx)^{h-1}B-{\mu^2(q^2+1)
\over q^2-1}h_{q^4}(xCx)^{h-1}+q^{4h}(xCx)^h\Delta;           \eqno (2.32)
$$
the operator $B$ is defined by
$$
B:=1+{q^2-1\over \mu}x^i\partial_i                           \eqno (2.33)
$$
and satisfies the properties $B(xCx)=q^2(xCx)B$, $B\Delta=q^{-2}\Delta B$.

{}~~~

{\bf 2.2. The Schroedinger equation for the harmonic oscillator.}

{}~

{}~~~In Ref. [14] we introduced the " hamiltonians "
$$
h_{\omega}:={1\over 2}(-q^N\Delta+\omega^2xCx)~~~~~~~~~\bar h_{\omega}:=
{1\over 2}( -q^{-N}\bar\Delta +{\omega}^2(xCx))                  \eqno (2.34)
$$
corresponding to the calculi $D,\bar D$. If $q=1$ both coincide with the
hamiltonian of the classical harmonic isotropic oscillator on ${\bf R}^N$.
The eigenvalues of $h_{\omega},\bar h_{\omega}$ coincide and
$$
h_{\omega}^*=\bar h_{\omega}.                                   \eqno (2.35)
$$
The eigenvalues $\{E_n\}_{n=1,2,..}$ are given by
$$
E_n=\omega(q^{{N\over2}-1}+q^{1-{N\over 2}})[{N\over 2}+n]_q ~~~~~~~~~~n \ge 0;
                                                                  \eqno (2.36)
$$
the corresponding $n^{th}$ level eigenfunctions are combinations of the
`` q-deformed Hermite functions '' $\psi_n^{i_n i_{n-1} ...i_1},\bar
\psi_n^{i_n i_{n-1} ...i_1}$
$$
h_{\omega}\psi_n^{i_n i_{n-1} ...i_1}=E_n\psi_n^{i_n i_{n-1} ...i_1}~~~~~~~~
\psi_n^{i_n i_{n-1} ...i_1}:=a^{i_n+}_na_{n-1}^{i_{n-1}+}...a_1^{i_1+}
\psi_0
$$
$$
\bar h_{\omega}\bar\psi_n^{i_n i_{n-1} ...i_1}=E_n\bar\psi_n^{i_n i_{n-1}
...i_1}~~~~~~~~
\bar\psi_n^{i_n i_{n-1} ...i_1}:=\bar a^{i_n+}_n\bar a_{n-1}^{i_{n-1}+}...
\bar a_1^{i_1+}\bar\psi_0,                                       \eqno (2.37)
$$
where the indices $i_j,~j=1,...,n,$ belong to $\{1,2,...,N\}$.
Here $\psi_0,\bar \psi_0$ denote the ground state eigenfunctions
$$
\psi_0:=exp_{q^2}[-{q^{-N}\omega xCx \over \mu}]~~~~~~~~~~
\bar\psi_0:=exp_{q^{-2}}[-{q^N\omega xCx \over \bar\mu}],        \eqno (2.38)
$$
and
$$
a_h^{i+}:=b_h(q)(x^i-{q^{2-h}\over \omega}\partial^i)G_q~~~~~~~~~~~
\bar a_h^{i+}:=b_h(q^{-1})(x^i-{q^{h-2}\over \omega}\bar\partial^i)G_{q^{-1}}
{}~~~~~~~~~i=1,2,...,N                                       \eqno (2.39)
$$
are the " creation " operators at level h in the unbarred and barred scheme
respectively. The operator $G_q$ was defined in (2.21) and
$$
[n]_q := {{q^n-q^{-n}}\over {q-q^{-1}}}                         \eqno (2.40)
$$
are the q-deformed integers: $[n]_q{{q \rightarrow 1}\atop \longrightarrow}n$.
At this stage we are free to fix the coefficients $b_h(q)$ as we wish.

The operators
$$
a_h^i:=d_h(q)(x^i+{q^{h+N}\over \omega}\partial^i)G_q~~~~~~~~~
\bar a_h^i:=d_h(q^{-1})(x^i+{q^{-h-N}\over \omega}\partial^i)G_{q^{-1}}
                                                               \eqno (2.41)
$$
are destruction operators (at level $h-2$), since $a_h^i\psi_{h-1},
\bar a_h^i\psi_{h-1}$ are eigenvectors of level $(h-2)$. Again, at this stage
we are free to fix the coefficients $d_n$ as we wish. In section 5
we will fix $b_n,d_n$ so as to build in the simplest way
well-defined position/momentum observables.

{}~~~As noticed in Ref. [14], the spectrum is bounded from below and
increasing
with n for any $q\in {\bf R}^+$. Energy levels are invariant under the
replacement $q\rightarrow q^{-1}$ and have the same degeneracy as in the
classical case. They are not equidistant as in the classical case ($q=1$) and
the difference between neighbouring energy levels diverges with $n$.

{}~~~In the classical case any function of the type $P(x)
exp[-{\omega (xCx)\over 2}]$ ($P(x)$ being a polynomial) can be expressed
as a combination of particular functions of this type, the well-known
Hermite functions with characteristic constant
$\omega$. Moreover, any eigenfunction of the corresponding harmonic oscillator
hamiltonian is a combination of Hermite functions of one (and the same) level.
Quite similar results hold also in the q-deformed case. The formulation of
these
results is however slightly technical; the reader who is not interested in
details can skip the rest of this section without relevant consequences for
the general understanding.

{}~~~According to the notation introduced in [14], let
$$
P_n(x):= a~polynomial~in~x~containing~only~powers~of~degree~p=n(mod~2).
                                                                \eqno (2.42)
$$
Using the q-derivative property (2.30) one checks that a function of the type
$polynomial\cdot gaussian$ can be expressed in infinitely many equivalent ways:
$$
P_n(x)exp_{q^2}[-{\omega q^{-N-m}(xCx)\over \mu}]=
$$
$$
\equiv P_{n+2h}(x)exp_{q^2}[-{\omega q^{-N-m-2h}(xCx)\over \mu}],
{}~~~~~~~~h \ge 0.                                               \eqno  (2.43)
$$
{}~~~Having defined the spaces $\Psi_n$, $V_n$, $V$ and $S$ by
$$
\Psi_n:=Span_{\bf C}\{\psi_n~'s~of~formula~(2.37)\}       \eqno (2.44)
$$
$$
V_n:=Span_{\bf C}\{ P_n(x)
exp_{q^2}[-{\omega q^{-n-N}(xCx) \over \mu}]\}.           \eqno (2.45)
$$
$$
V:=\sum\limits_{n=0}^{\infty} V_n                         \eqno(2.46)
$$
and
$$
S:=Span_{\bf C}\{eigenfunctions~\psi~of~h_{\omega}~of~the~form~
\psi=P(x)exp_{q^2}[-{\alpha(xCx)\over \mu}],                  \eqno (2.47)
$$
then one can prove

{\bf Proposition 2.1:}
$$
V=S=\bigoplus\limits_{n=0}^{\infty}\Psi_n.                          \eqno
(2.48)
$$
This equality has the following

{}~~

{\bf Corollary 2.2}:
No eigenfunctions of $h_{\omega}$ other than the ones belonging to the
$\Psi_n$'s can be found in $V=S$. Correspondingly, no eigenvalues other than
$E_n$'s $n=1,2,...$.

{}~~~For the proof [14] of statement (2.48) one makes essential use of relation
(2.43) and of the property
$$
{\cal P}_{A~hk}^{~~ij}a_n^{h+}a_{n-1}^{k+}=0                \eqno (2.49)
$$
and proves the

{\bf Lemma 2.3}
$$
V_n=\bigoplus \limits_{0\le h \le {n\over 2}}\Psi_{n-2h};
{}~~~~~~~~~~dim(\Psi_n)=dim(M_n)={N+n-1 \choose N-1},
{}~~~~~~~~~~~~~n \in N.                                             \eqno
(2.50)
$$
Here
$$
M_n:=Span_{\bf C}\{x^{i_1}x^{i_2}...x^{i_n}, i_h=1,2,...N\}.       \eqno (2.51)
$$

{}~~~Note that $V=\bigoplus \limits_{n=0}^{\infty} \Psi_n$. $V$ can be split
into
subspaces $V^+,V^-$ of opposite parity:
$$
V^+:=\bigoplus \limits_{h=0}^{\infty} \Psi_{2h}~~~~~~~
V^-:=\bigoplus \limits_{h=0}^{\infty} \Psi_{2h+1},~~~~\Rightarrow
{}~~~~V=V^+\oplus V^-                                              \eqno (2.52)
$$

{}~~

{}~

\centerline{\bf {\large 3.} Integration: formal requirements}

{}~~

{}~~~In Ref. [7] the authors propose a definition of integration
over the quantum hyperplane essentially based on the requirements of
linearity and of validity of Stoke's theorem (of course in such an approach
the latter is no more a `` theorem ''). Denoting by $<f>$, $\int \omega_n$
respectively the integral of a function $f$ and of an n-form $\omega_n$
over the n-dimensional hyperplane (as usual they are related by definition
by the identity $<f>:=\int dVf$, where $dV$ denotes the volume form),
Stoke's theorem takes respectively the forms
$$
<\partial^i f|>=0~~~~i=1,2,...,n,~~~~~~~~~~~~~~~\int d\omega_{n-1}|=0;
                                                                    \eqno (3.1)
$$
$\partial^i f|$ denotes the (total derivative) function which was introduced in
formula (2.27).
In the classical case, if $f=P_n(x)exp[-a|x|^2]$
($P_n$ denotes a polynomial of degree $n$ in $x$ and $|x|^2$ the square
lenght), then
$$
\partial^i P_n(x)exp[-a|x|^2]|=  P_{n-1}(x)exp[-a|x|^2]+
 P_{n+1}exp[-a|x|^2];                                           \eqno (3.2)
$$
relations (3.1), (3.2) imply
$$
<  P_{n-1}(x)exp[-a|x|^2]>+ < P_{n+1}exp[-a|x|^2]>=0.         \eqno (3.3)
$$
Relation (3.3) allows to recursively define
the integral $<f>$ (for any function $f$ of the same kind)
in terms of $<exp[-a|x|^2]>$
(which fixes the normalization of the integration). The same holds in the
q-deformed case, provided one has defined the generalization of the exponential
(the socalled q-exponential).

{}~~~The integration over the hyperplane defined according to (3.1), (3.2) has
the
following properties: a) it is continuous in $q$; b) it is
covariant w.r.t. $GL_q(n)$ (in the sense that will defined below);
c) it coincides with the classical Riemann integral for q=1
(by a suitable choice of the normalization factor); d) it satisfies
the reality condition
$$
<f>^*=<f^*>                                                      \eqno (3.4)
$$
for any q ($\in {\bf R}^+$); therefore: d) the positivity condition,
$$
<f^*f> \ge 0,~~~~~~~~~~~~~~<f^*f>=0 \Leftrightarrow f=0,          \eqno (3.5)
$$
holds, at least in a ($f$-dependent) neighbourhood of
q=1, since it holds for q=1. If there exists a neighbourhood
$U\subset {\bf R}^+$ of q=1 such that positivity holds $\forall q\in U$,
in this neighbourhood a scalar product can be introduced through the definition
$$
(f,g):=<f^*g>,                                                    \eqno (3.6)
$$
and one can convert into a Hilbert space a suitable subspace of the algebra
of functions on the quantum hyperplane.

{}~~~In the case of the real quantum euclidean space the situation seems
complicated by the fact that there exist two sets of linearly
independent derivatives
belonging respectively to the differential calculi $D,\bar D$, hence
potentially two kinds
of integrations $<~~>,\ll~~\gg$ and two versions of Stoke's theorem:
$$
<\partial^i f|>=0~~~~i=1,2,...,N;~~~~~~~~~~~~~~\int d\omega_{n-1}|=0
$$
$$
\ll\bar\partial^i f|\gg=0~~~~i=1,2,...,N;~~~~~~~~~~~~~~\bar{\int}\bar d \bar
\omega_{n-1}|=0.                                                   \eqno (3.7)
$$
At first sight the reality condition (3.4) for each of the
two integrations $<~~>,\ll~~\gg$ seems no more guaranteed by Stoke's theorems
(3.7) because $*$ maps derivatives $\partial\in D$ into derivatives
$\bar \partial\in \bar D$ (and viceversa). Quite surprisingly, in next section
we will see that the two integrations coincide; therefore relation (3.4)
holds.

{}~~~For the moment we keep $<~~>$ and $\ll~~\gg$ distinct. We list the
requirements that these integrations should satisfy and show that they are
compatible with each other. In next section one of these requirements,
Stoke's theorem, will be used to recursively define the integrations.
Through the relation
$$
<f>=\int dV~f~~~~~~~~~~                                          \eqno (3.8)
$$
statements
regarding integral of functions can be translated into ones regarding integrals
of $N$-forms, and viceversa, so usually they will be written only in one of the
two versions.

{}~~~We would like an integration $<~~>$ to be defined on a not too poor
subspace ${\cal V}$  of $Fun({\bf R_q^N})$ and to satisfy:

1) linearity;

2) covariance;

3) continuity in $q$ and correspondence principle for $q\rightarrow 1$;

4) reality;

5) positivity.

{}~~~Of course linearity means
$$
<\alpha f+\beta g>=\alpha <f>+\beta <g>,~~~~~~~\alpha,\beta \in {\bf C}
{}~~~f,g\in {\cal V}                                             \eqno (3.9)
$$
and one has to check that if $f$ vanishes because of relations (2.9), then
so does $<f>$, in other terms
$$
f(x)=A_{ij}{\cal P}^{~~ij}_{A~hk}x^hx^k \cdot g(x)~~~~\Rightarrow
{}~~~<f>=0.                                                     \eqno (3.10)
$$

{}~~~By covariance we mean
$$
{\bf 1}_{SO_q(N)}<f>=(id_{SO_q(N)}\bigotimes <~~>) \circ \phi_L(f),
                                                              \eqno (3.11)
$$
where ${\bf 1}_{SO_q(N)}$ and $id_{SO_q(N)}$ denote respectively the unit
element and the identity operator on $Fun(SO_q(N,{\bf R}))$, and $\phi_L$ is
the left coaction of $SO_q(N,{\bf R})$ on $Fun({\bf R_q^N})$. More explicitly,
if $f^{i_1i_2...i_k}:=x^{i_1}x^{i_2}...x^{i_k}g(xCx)$, then covariance
means
$$
{\bf 1}_{SO_q(N)}<f^{i_1i_2...i_k}>=T^{i_1}_{j_1}T^{i_2}_{j_2}...T^{i_k}_{j_k}
<f^{j_1j_2...j_k}>,                                         \eqno (3.12)
$$
in other words the numbers $<f^{i_1i_2...i_k}>,~~i_j=1,2,...,N$, are the
components of an " isotropic " tensor; in the classical case
relation (3.12) corresponds to the well-known property of tensors such as
$$
\int d^Nx~g(|x|^2)x^i=0,~~~~~~~~\int d^Nx ~g(|x|^2) x^ix^j \propto \delta^{ij},
$$
$$
\int d^Nx~g(|x|^2)x^ix^jx^kx^l \propto (\delta^{ij}\delta^{kl}+
\delta^{ik}\delta^{jl}+\delta^{il}\delta^{jk}),...           \eqno (3.13)
$$
namely the property that the latter
are invariant under an orthogonal transformation of the
coordinates $x^i\rightarrow x^{'i}:=g^i_jx^j$. The simplest nontrivial
example of a tensor satisfying (3.12) is for $k=2$, $<f^{ij}>\propto C^{ij}$.
In general tensors satisfying (3.12) involve matrix products among $\hat
R$-matrices (or, equivalently, $\hat R^{-1}$-matrices) and contractions
with metric matrices $C$: the former reorder indices by means
of the RTT relations (2.2),  whereas the latter transform a couple of
neighbouring $T$-matrices into a commuting number (see relation (2.1)).
Therefore an integral
$<x^{i_1}...x^{i_k}g(xCx)>$ should be factorizable as a product
$$
<x^{i_1}...x^{i_k}g(xCx)>= S^{i_1...i_k} \alpha_g,                \eqno (3.14)
$$
and $S^{i_1...i_k}=0$ for $k$ odd; the $g$-dependence of the RHS of (3.14) is
concentrated in the constant $\alpha_g$, which essentially is a (yet
unspecified) integral along the `` radial '' direction. Explicit solutions
$S^{i_1...i_k},\bar S^{i_1...i_k}$ satisfying (3.12) will be found in
setion 4.

{}~~~Point 3) means that we require a q-deformed integral to reduce to a
classical one (with some integration measure $\rho(x)d^Nx$) when q=1.
Maybe it is timely to  recall the fact that the $x^i$ coordinates are not
real (even for q=1), but are complex combinations of the usual real
cartesian coordinates; the latter can be used to perform the integral when
q=1.

{}~~~The reality and positivity conditions 4), 5) in the form (3.4),(3.5) or in
some other form should guarantee that the definition (3.6) or what takes
its place introduces an honest scalar product (~~,~~) in a suitable subspace
${\cal V}$ of $Fun({\bf R_q^N})$
$$
(f,g)^*=(g,f);~~~~~~~~~~(f,f)\ge 0,~~(f,f)=0 \Leftrightarrow f=0
 ~~~~~~~f,g \in {\cal V},                                        \eqno (3.15)
$$
to convert this subspace into a Hilbert space.

{}~~~To the five previous points we add  a requirement characterizing
the specific problem we are dealing with here, namely that the hamiltonian
(or the position/momentum operators) of
the harmonic oscillator be hermitean operators w.r.t. $(~~,~~)$. As it will be
clear in the sequel, we are led to ask for the validity of

{}~~

6) Stoke's theorem

{}~~

in the form (3.7). We will see that, as in the classical case, point 6)
involves a definite
choice of the " radial " part of the integration (whereas the latter is left
unspecified by 2) alone). As already noticed, Stoke's theorem is a formidable
tool to define (up to a normalization factor) the corresponding integration.

{}~~~Now we briefly discuss compatibility of requirements 1) - 6).

{}~~~It is straightforward to check that linearity is compatible with
covariance because of property (2.6) (where we take $f(\hat R)={\cal P}_A$).
Requirement 3) is obviously compatible with 1),2) since classical integration
is linear and there exist $SO(N,{\bf R})$-invariant integration measures.
 It is straightforward to prove that reality (3.4)
is compatible with linearity (because of relation (2.6), where again
we take $f(\hat R)={\cal P}_A$), with covariance (apply
$*$ to eq. (3.12) and use definitions (2.23),(2.24)) and with the
correspondence
principle (classical real integrations satisfy the reality condition).
Positivity in the form (3.5) is clearly compatible with requirements
1),3) and with reality (3.4). At this stage is
not easy to understand if it is compatible with covariance. Using the
results which will be presented in Sect.'s 6.,7. one could prove that this
is the case. The question
is not strictly relevant for the solution of problem (3.15)
in the specific case of the harmonic oscillator, since the scalar
product $(~~,~~)$ that we are going to introduce in Sect. 5 is not of the form
(3.6). In fact, we will prove that the scalar product introduced there
 $is$ positive definite.

{}~~~Let us analyze now the compatibility of Stoke's theorems (3.7) with points
1) - 5). It is straightforward to check that both $(3.7)_a$ and $(3.7)_b$
are compatible with linearity: in fact this compatibility is reduced to the
consistency of both differential calculi $D,\bar D$ with the q-space
relations (2.9) and the commutation relations (2.13).
Similarly, the compatibility with covariance is guaranteed by the fact that
the exterior derivative commutes with the coaction (see formula $(2.17)_a$).
As for the
correspondence principle, compatibility is ensured by the fact that
in the limit $q\rightarrow 1$ both
$D$ and $\bar D$ go to the classical differential calculus, and Riemann
integration (on smooth functions) satisfies Stoke's theorem. In order to
show that Stoke's theorems are compatible with reality in the form (3.4),
we first show that

{}~

{\bf Proposition 3.1}: the integrations $<~~>,\ll~~\gg$ satisfying Stoke's
theorems $(3.7)_a$ and $(3.7)_b$ are compatible with reality in the form
$$
<f>^*=\ll f^*\gg,                                         \eqno (3.16)
$$

{}~~~Then compatibility with reality in the form (3.4)
will follow from the equality $<~~>=\ll~~\gg$, which we will prove in next
section.

{}~~~$Proof$: Let us consider the spaces of formal relations
$$
{\cal F}=Span_{\bf C}\{\partial^i f-\partial^i f|-f^i_j\partial^j=0,
{}~~~~~~~i,j=1,..,N~~~~f\in {\cal V}\}
                                                              \eqno (3.17)
$$
$$
\bar{\cal F}:=Span_{\bf C}\{\bar\partial^i f-\bar\partial^i f|-\bar f^i_j\bar
\partial^j=0,~~~~~~~i,j=1,..,N~~~~f\in {\cal V}\},            \eqno (3.18)
$$
where: 1) $\partial^i f|,f^i_j,\bar\partial^i f|,\bar f^i_j$ are the functions
introduced in formula (2.27); 2) ${\cal V}$ is some subspace of
$Fun({\bf R_q^N})$ closed under complex conjugation and containing also the
functions $\partial^i f|,f^i_j,\bar\partial^i f|,\bar f^i_j$ for any
$f\in {\cal V}$. In the classical
case the space $V_{cl}$ of functions of the type $P(x)exp[-a|x|^2]$ ($P$ being
a polynomial) is an example of such a subspace ${\cal V}$, and we will see
that an analogous space will be available in the q-deformed case, too.
Under these assumptions  it is immediate to recognize that the two sets
(3.17),(3.18) are mapped into each other by $*$, since $*:D\rightarrow \bar D$
and $*:\bar D\rightarrow D$. In other terms ${\cal F}^*=\bar {\cal F}$. If
we define subspaces ${\cal A},\bar {\cal A}\subset {\cal V}$ as the linear
spans of functions $\partial^if|$ and $\bar \partial^i f|$ respectively, the
previous remark implies
$$
{\cal A}^*=\bar {\cal A}                                       \eqno (3.19)
$$
For each $a\in {\cal A}$ let $\bar a\in \bar {\cal A}$ be the function such
that
$a^*=\bar a$. Stoke's theorems respectively imply
$$
(3.7)_a ~~~~\Rightarrow ~~~~~~~~<a>=0=<a>^*~~~~\forall a\in {\cal A}
                                                              \eqno (3.20)
$$
$$
(3.7)_b ~~~~\Rightarrow ~~~~~~~~\ll \bar a\gg=0=\ll\bar a\gg^*~~~~
\forall \bar a \in \bar{\cal A},                              \eqno (3.21)
$$
hence reality in both forms (3.4) and (3.16) is trivially satisfied for the
integrals $<a>$, $\ll \bar a \gg$. If q=1 and we take ${\cal V}=V_{cl}$ one
easily realizes that any $f\in {\cal V}$ can be expressed in the form
$$
f=a+c_ff_0,~~a\in {\cal A},~c_f\in {\bf C}                  \eqno (3.22)
$$
(as anticipated at the beginning of this section), where $f_0$ is defined by
$f_0:= exp[-a|x|^2]$. Consequently
$$
<f>=c_f<f_0>.                                                 \eqno (3.23)
$$
For self-evident reasons we call $f_0$ the reference function of the
integral.
In next sections we will see that a similar situation occurs also in the
q-deformed case, for instance by taking ${\cal V}=V$ ($V$ was defined in
Sect. 2.) and $f_0:=exp_{q^2}[-a{(xCx)\over \mu}]$. In any case $f_0$ should be
a real function not belonging to ${\cal A}$ and should go to a smooth
rapidly decreasing classical function in the limit $q\rightarrow 1$.
Taking the complex conjugate of eq. (3.22) we get
$$
f^*=\bar a+c_f^*f_0,~~\bar a\in \bar{\cal A},~c_f\in {\bf C},     \eqno (3.24)
$$
which implies
$$
\ll f^* \gg=c_f^* \ll f_0\gg                                      \eqno (3.25)
$$
We are still free to fix $<f_0>,\ll f_0\gg$ as we like. If we impose the
reality
condition in the form (3.16) on the reference function we see that
it is transfered to all functions belonging to ${\cal V}$, as claimed
$\diamondsuit$.

{}~

{}~~~Since $f_0^*=f_0$, the reality condition (3.16) on the reference function
reads $<f_0>=\ll f_0 \gg^*$. In the sequel we will take $<f_0>\in {\bf R}^+$.

{}~~~Finally the compatibility of Stoke's theorem with the positivity condition
in the form (3.5) is left as an open question, but again is not relevant for
our specific problem; whereas we will see in Sect. 7 that the scalar product
in the Hilbert space of the harmonic oscillator, defined using
the integral $<~~>$, is positive defined.

{}~~

{}~~

\centerline{\bf {\large 4}. Integration: construction}

{}~~

{}~~~In this section we use Stoke's theorem (in its two
versions (3.7)) as a tool for constructing the integrations. The systematic
enforcement
of Stoke's theorems generates a set of formal relations between integrals
of different functions. We determine these relations in two steps. First,
we find out the isotropic tensors $S^{i_1...i_k}, \bar S^{i_1...i_k}$: hence,
according to (3.14) , the integrals $<f>,~\ll f \gg$ of a non scalar function
$f$ will be expressed in terms of integrals of a scalar one. Due to the fact
that $S^{i_1...i_k}, \bar S^{i_1...i_k}$ turn out to be proportional, the
two integrations coincide. Second, we
determine the equations relating integrals of different scalar functions;
in this way we will be able to express integrals of scalar
functions in terms of the integral $<f_0>$ of a particular
one, what we call the reference function $f_0$. $<f_0>$ is a
normalization constant and can be fixed quite
arbitrarily (see the end of the preceding section). So to say, the second
step amounts to integration over the radial coordinate. As an example we will
explicitly consider in this section the reference function
$f_0=exp_{q^2}[{-\alpha xCx \over \mu}]$; in section 5. we will take
an other reference function which is conceived for defining the scalar
products of states of the harmonic oscillator. In this way one can define
the integrals for infinitely many independent functions $\{f_i\}_{i\in N}$ and
therefore for finite combinations of them. This is enough for the scopes of
this work, since it will enable us to define a positive definite scalar
product inside the span of states of the harmonic oscillator (see section 5.);
then the completion of this pre-Hilbert space will be done w.r.t. the
corresponding norm. Nevertheless, to further enlarge the domain of
definition of the integrals one could consider functions admitting
series expansions in the $\{
f_i\}$, and we will briefly address this problem at the end of this section.

{}~~~The preliminary
discussion of the previous section has shown that the two basic integrations
$<~~>,~~\ll~~\gg$ are linear, covariant and coincide with the classical
Riemann integration for q=1. Therefore the explicit recursive
application of the two Stoke's theorems will determine (up to a factor)
isotropic tensors $S^{i_1...i_k}$, $\bar S^{i_1...i_k}$ (see (3.12)).
As we are going to see, up to a factor these tensors coincide and
do not depend on the choice of the function $g(xCx)$ in formula (3.14).
The relevant results of this section are summarized in Propositions
4.1, 4.2, 4.3.

{}~~~The choice $g=exp_{q^2}[{-\alpha xCx \over \mu}]$ (or, alternatively, we
could take
$g=exp_{q^{-2}}[{-\alpha xCx \over \bar\mu}]$) is particularly convenient
for this goal. Using relation (2.29)  we find
$$
\partial^{i_1}x^{i_2}...x^{i_k}exp_{q^2}[{-\alpha xCx \over \mu}]|=
$$
$$
=-\alpha x^{i_1}x^{i_2}...x^{i_k}exp_{q^2}[{-\alpha xCx \over \mu}]+
exp_{q^2}[{-q^2\alpha xCx \over \mu}]\partial^{i_1}x^{i_2}...x^{i_k}|=
$$
$$
=-\alpha x^{i_1}x^{i_2}...x^{i_k}exp_{q^2}[{-\alpha xCx \over \mu}]+
exp_{q^2}[{-q^2\alpha xCx \over \mu}]M^{~~i_1...i_k}_{k,j_3...j_k}
x^{j_3}...x^{j_k}                                              \eqno (4.1)
$$
where the tensors $M^{~~i_1...i_k}_{k,j_3...j_k}$,
$N^{~~i_1...i_k}_{k,j_1...j_k}$ are introduced by the defining relation
$$
\partial^{i_1}x^{i_2}...x^{i_k}=:M^{~~i_1...i_k}_{k,j_3...j_k}
x^{j_3}...x^{j_k}+N^{~~i_1...i_k}_{k,j_1...j_k}x^{j_1}...x^{j_{k-1}}
\partial^{j_k}                                                  \eqno (4.2)
$$
Taking the integral $<~~>$ of (4.1) and applying Stoke's theorem we find
$$
<x^{i_1}...x^{i_k}exp_{q^2}[-{\alpha xCx \over \mu}]>=
{1\over \alpha}M^{~~i_1...i_k}_{k,j_3...j_k}
<x^{j_3}...x^{j_k}exp_{q^2}[-{q^2\alpha xCx \over \mu}]>        \eqno (4.3)
$$
Starting from $k=0,1$ and noting that Stoke's theorem
(or, equivalently, covariance) imply $<x^iexp_{q^2}[-{\alpha xCx \over
\mu}]>=0$, we see that the recursive application of relation (4.3)
determines tensors $S_{k}^{i_1...i_k}$. The result is summarized in the

{}~

{\bf Proposition 4.1}: the tensors
$$
S_{k}^{i_1...i_k}:=0~~if~k~is~odd                                 \eqno (4.4)
$$
$$
S_{2n}:=M_{2n}\cdot M_{2(n-1)}\cdot...M_2                         \eqno (4.5)
$$
satisfy the covariance condition (3.12).

Here we have used the shorthand notation
$$
(M_{2k}\cdot M_{2(k-1)})^{i_1i_2...i_{2k}}_{j_5j_6...j_{2k}}=
M^{~~~i_1i_2...i_{2k}}_{2k,l_3l_4...l_{2k}}M^{~~~~~~~l_3l_4...l_{2k}}
_{2(k-1),j_5j_6...j_{2k}}.                                        \eqno (4.6)
$$
As a direct consequence of the proposition and of relation (4.3),
the integral (4.3) will vanish when $k$ is odd and
$$
<x^{i_1}x^{i_2}...x^{i_{2n}}exp_{q^2}[-{q^2\alpha xCx \over \mu}]>
\propto S_{2n}^{i_1i_2...i_{2n}}.                                 \eqno (4.7)
$$
Similarly one can determine tensors $\bar S_{k}^{i_1...i_k}$ satisfying the
analog of relation (3.12) for the integration $\ll ~~\gg$; to this end we
only need to replace $\partial,M,N,S,\hat R^{\pm}$ with
$\bar\partial,\bar M,\bar N,\bar S,\hat R^{\mp}$
in the preceding  formulae.

{}~

{\bf Proposition 4.2}:
$$
\bar S_{2n}^{i_1...i_{2n}}\propto S_{2n}^{i_1...i_{2n}}.         \eqno (4.8)
$$
$Proof$: this immediately follows from the very useful formulae
$$
\Delta^n x^{i_1}x^{i_2}...x^{i_2n}|=(\mu)^n n_{q^2}!S_{2n}^{i_1i_2...i_{2n}}
                                                                  \eqno (4.9)
$$
$$
\bar\Delta^n x^{i_1}x^{i_2}...x^{i_2n}|=(\bar\mu)^n n_{q^{-2}}!\bar S_{2n}^
{i_1i_2...i_{2n}}                                               \eqno (4.10)
$$
and from equations (2.20),(2.21). The proof of relations (4.9),(4.10)
is by induction and will be given in appendix A $\diamondsuit$.

{}~~~Next, it is easy to realize that equation (4.7) can be generalized to
$$
<x^{i_1}x^{i_2}...x^{i_{2n}}g(xCx)>
\propto S_{2n}^{i_1i_2...i_{2n}}.                                 \eqno (4.11)
$$
with any function $g(xCx)$. In fact, looking
at the power series defining $g$ one immediately finds that $\partial^i
g(xCx)|=\tilde g(xCx)x^i$, with some functions
$\tilde g,\in Fun({\bf R_q^N})$. Then, applying
both sides of (4.2) to $g$ and taking the integral $<~~>$ we find
$$
0=<\partial^{i_1}x^{i_2}...x^{i_2n}g|>=M^{~~~i_1i_2...i_{2n}}_{2n,j_3...
j_{2n}}<x^{j_3}...x^{j_2n}g>+N_{2n,j_1...j_{2n}}^{~~~i_1...i_{2n}}<x^{j_1}...
x^{j_{2n}}\tilde g>.                                           \eqno (4.12)
$$
This result holds for any function $g$, in particular for the
previous choice $g=exp_{q^2}[{-\alpha xCx \over \mu}]$; by comparison with
(4.3), (4.5), we infer the invertibility of the matrices $N_{2n}$,
the relations
$$
N_{2n}^{-1}\cdot S_{2n}\propto S_{2n}                       \eqno (4.13)
$$
and hence the relations
$$
<x^{i_1}...x^{i_{2n}}\tilde g>=c_{n,g} S_{2n}^{i_1...i_{2n}},    \eqno (4.14)
$$
for any function $\tilde g(xCx)$. By contracting the free indices
$i_1,i_2,...,i_{2n}$ with $C_{i_1i_2}$,...,$C_{i_{2n-1}i_{2n}}$ we reduce the
determination of the constant $c_{n,\tilde g}$ to the
evaluation of the integral of a purely scalar function. The same
arguments can be applied to the integration $\ll~~\gg$. Thus we are led to the

{}~

{\bf Proposition 4.3:}
$$
<x^{i_1}...x^{i_{2n}}\tilde g>=S_{2n}^{i_1...i_{2n}}{<(xCx)^n \tilde g> \over
{\cal S}_{2n}},                                                  \eqno (4.15)
$$
$$
\ll x^{i_1}...x^{i_{2n}}\tilde g\gg=S_{2n}^{i_1...i_{2n}}{\ll
(xCx)^n \tilde g\gg \over{\cal S}_{2n}};                 \eqno (4.16)
$$
here
$$
{\cal S}_{2n}:=C_{i_1i_2}...C_{i_{2n-1}i_{2n}} S_{2n}^{~~i_1,i_2,...,i_{2n}}.
                                                                \eqno (4.17)
$$
{}~~~Using formulae (2.32),(4.9) it is easy to show that the costant
${\cal S}_{2n}$ is positive for any $q\in {\bf R^+}$.

{}~

{}~~~Let us analyze the `` radial '' dependence of the two integrals $<~~>$,
$\ll ~~ \gg$. We introduce the operators
$$
B:=1+{q^2-1\over \mu}x^i\partial_i=q^{-N}(1+{q^2-1\over \mu}\partial^ix_i)
$$
$$
\bar B:=1+{q^{-2}-1\over \bar\mu}x^i\bar\partial_i=q^N(1+{q^{-2}-1\over \bar
\mu}\bar\partial^ix_i);                                        \eqno (4.18)
$$
it is straightforward to check that $B(xCx)=q^2(xCx)B$, $\bar B(xCx)=q^{-2}
(xCx)\bar B$ and therefore
$$
Bf(xCx)=f(q^2xCx)B,~~~~~~\bar Bf(xCx)=f(q^{-2}xCx)\bar B,       \eqno (4.19)
$$
for any $f \in O_q(N)$ depending only on $(xCx)$; hence
$$
q^{-N}(f+{q^2-1\over \mu}\partial^ix_i f|)=f(q^2xCx),~~~~~~~~~~~~
q^N(f+{q^{-2}-1\over \bar\mu}\bar\partial^ix_i f|)=f(q^{-2}xCx).   \eqno (4.20)
$$
By taking the integrals $<~~>$, $\ll~~\gg$ respectively of $(4.20)_a$,
$(4.20)_b$ and by applying Stoke's theorems (3.7) we find the formal relations
$$
<f(q^2xCx)>q^N=<f(xCx)>,~~~~~~~~~~\ll f(q^2xCx)\gg q^N= \ll f(xCx) \gg
                                                                  \eqno (4.21)
$$
for both integrations. As we will see in a moment, equation (4.21) is
sufficient to determine the integral of any scalar function in terms of that
of the reference function $f_0$; if we set
$<f_0>=\ll f_0\gg(\in {\bf R^+})$, this implies the formal relation
$$
<f>=\ll f\gg,                                                      \eqno (4.22)
$$
at least for $f=f(xCx)$. But looking back at relations (4.16),(4.8) we realize
that previous equation holds for any $f$. This concludes the proof of the

{}~

{\bf Proposition 4.4:} the two integrations $<~~>,\ll~~\gg$ (formally)
coincide.

{}~

{}~~~Since the integral $<f>$ of any $f\in Fun({\bf R_q^N})$,
if it exists, is reduced to a
combinations of integrals of radial functions by means of  relation (4.15),
then property (4.21) is generalized by the

{}~

{\bf Proposition 4.5:}
$$
<f(qx)>q^N=<f(x)>.                                              \eqno (4.23)
$$
This fundamental relation characterizes the integration
defined by means of Stoke's theorem and will be called " scaling property "
for reasons which
will become clear at the end of this section.

{}~~~So far we have not specified the domain of functions
$f\in Fun({\bf R_q^N})$ for which the integral $<f>$ can be defined.
Therefore all the previous relations were purely formal. Now we pick up
a particular reference function $f_0(xCx)$. We ask what are the functions $f$
such that the corresponding integral $<f>$ can be reduced to the
one $<f_0>$ by means of iterated application of Stoke's theorem
and of linearity, and turn out to be finite.
Of course we wish to include in this space of " integrable "
functions as many $f\in Fun({\bf R_q^N})$ as possible.

{}~~~As an example we take $f_0=exp_{q^2}[-{\alpha xCx \over\mu}],~~\alpha>0$,
which for q=1 reduces to a well known
smooth rapidly decreasing classical function, the gaussian. First we
consider functions $f$ of the type $f(xCx)= P(xCx)exp_{q^2}[-{\alpha xCx \over
\mu}]$, $P$ being an arbitrary polynomial. Using property (4.23) and the
q-derivative property (2.30) of the exponential we show the

{\bf Proposition 4.6:}
$$
<exp_{q^2}[-{\alpha xCx \over \mu}](xCx)^h>=\left ({\mu \over \alpha}\right )^h
\left (h-1+{N\over 2}\right )_{q^2}...\left ({N\over 2}\right )_{q^2}\cdot
q^{-h(N+h-1)}c,          \eqno (4.24)
$$
where $c:=<exp_{q^2}[-{\alpha xCx \over \mu}]>\in {\bf R^+}$ plays the role
of normalization factor.

$Proof:$
$$
<exp_{q^2}[-{\alpha xCx\over \mu}](xCx)^{k-1}>=q^{N+2(k-1)}
<exp_{q^2}[-{q^2\alpha xCx \over\mu}](xCx)^{k-1}>=
$$
$$
=q^{N+2(k-1)}<exp_{q^2}[-{\alpha xCx \over\mu}](xCx)^{k-1}>+
$$
$$
-{\alpha \over  \mu}(q^2-1)q^{N+2(k-1)}
<exp_{q^2}[-{\alpha xCx \over\mu}](xCx)^k>,                  \eqno (4.25)
$$
whence
$$
<exp_{q^2}[-{\alpha xCx \over \mu}](xCx)^k>=({\mu \over \alpha})
(k-1+{N\over 2})_{q^2}q^{-N-2(k-1)}
<exp_{q^2}[-{\alpha xCx \over \mu}](xCx)^{k-1}>;                 \eqno (4.26)
$$
applying $h$ times formula (4.26) for $k=h,h-1,...,1$ we find (4.24)
$\diamondsuit$.

{}~~~Relations (4.15), (4.24) allow to define the
integration $<~~>$,  on all functions of the type
$f=P(x)f_0$, where $P(x)$ is an arbitary polynomial in $x$ and
$f_0:=exp_{q^2}[-{\alpha xCx \over \mu}]$. We could enlarge
the domain of definition of the integrations by admitting functions
$P(x)$ in the form
of power series $P(x)=\sum \limits_{n=0}^{\infty}\sum \limits_{i_1,i_2,...,i_n}
A_{i_1i_2...i_n}x^{i_1}x^{i_2}...x^{i_n}$ such that the series
$$
\sum \limits_{n=0}^{\infty}\sum \limits_{i_1,i_2,...,i_n}
A_{i_1i_2...i_n}<x^{i_1}x^{i_2}...x^{i_n}f_0>                   \eqno (4.27)
$$
converges;
the integral $<f>$ would then be defined as the limit
$(4.27)$. A further step towards the
enlargement of the domain of definition of the integrations could be done along
the following lines. In the previous formula we could take $P=P(xCx)=
\sum \limits_{n=0}^{\infty}a_n (xCx)^n$ with coefficients $a_n \in {\bf R}$ and
such that $<P(xCx)exp_{q^2}[-{\alpha xCx \over \mu}]>:=c'$ is finite.
Then we could define a new reference function by letting $f_0:=
P(xCx)exp_{q^2}[-{\alpha xCx \over \mu}]$: by means of formulae (4.15), (4.16),
(4.21) we should be able to evaluate $<\tilde P(x)f_0>$
in terms of $c'$ for all polynomials $\tilde P(x)$.
Thus one could include in the domain of integrable functions also functions
$f$ susceptible of a decomposition $f=\tilde P(x)f_0$, $\tilde P(x)=
\sum \limits_{n=0}^{\infty}
\sum \limits_{i_1,i_2,...,i_n}A_{i_1i_2...i_n}x^{i_1}x^{i_2}...x^{i_n}$ such
that the series (4.27) with this new $f_0$ converges.
It is natural to figure that to
the new choice of the reference function there should correspond an actual
enlargement of the domain of integrable functions.
This operation could be iterated in a sort of continuation of
the functional $<~~>$ so as to enlarge to the maximum possible
size the space of integrable functions.
It is out of the scope of this work to face this problem
by analyzing which conditions the coefficients $\{A_{i_1i_2...i_n}\}$
of an expansion of the type $f=f_0\sum \limits_{n=0}^{\infty}\sum \limits_
{i_1,i_2,...,i_n}A_{i_1i_2...i_n}x^{i_1}x^{i_2}...x^{i_n}$ should satisfy
in order that $f$ be integrable$^{(1)}$.

{}~~~We just briefly note that, having defined the integrations
$<~~>$ using Stoke's theorems (3.7), one could define a new integration
$<~~>_{\rho}$satisfying (at least) requirements 1) - 3)
of the preceding section, by setting
$$
<f>_{\rho}:=<f \cdot \rho>;                                  \eqno (4.28)
$$
the " weight " $\rho$ should be a real scalar function.

{}~~~Now let us come back to property (4.23). By its iterative application we
find
$$
<f(q^nx)>q^{nN}=<f(x)>,
{}~~~~~~~~~~~~~~~~~n\in {\bf Z},                                  \eqno (4.29)
$$
or, equivalently, in differential form notation
$$
\int dV~q^{nN}f(q^nx)=\int dV~ f(x),
{}~~~~~~~~~~~~n\in {\bf Z}.                                  \eqno (4.30)
$$
Relation (4.30) states that under the change of integration variables
$x\rightarrow ax$ with $a=q^n$ the integral $\int$ is invariant
if we let $dV$ transform according to $dV\rightarrow a^NdV$,
 namely according to the law of transformation of $d^Nx$. This
explains the name " scaling property " for relations (4.21),(4.23),(4.29),
(4.30).
In both the classical and the q-deformed case this property characterizes
the integrals satisfying Stoke's theorem; for q=1 the latter reduce to the
usual Riemann integral, which has a " homogeneous " (i.e. translation
invariant)
measure.

{}~~~One can now ask if the scaling property holds even if the
dilatation parameter $a \notin Q:=\{q^n,~n\in {\bf Z}\}$. One can easily
check that this is not the case (let for instance $f_0=exp_{q^2}[-b(xCx)]$
and let $f=exp_{q^2}[-ab(xCx)]$  be the function which we want to
integrate by choosing $f_0$ as reference function).
In other terms, the function
$$
F(a):=<f(ax)>a^N                                               \eqno (4.31)
$$
is periodic in the variable $b=ln(a)$ with period $ln(q)$, but is not
constant. To be specific, assume for instance $q>1$.
If $a\in {\bf R}^+$ is fixed and $q^n<a<q^{n+1}$ the function $F$ fluctuates
around the value $F(1)=<f(x)>$, the width of the fluctuation being
the same $\forall n\in {\bf Z}$, therefore also around large $a$.
But if we take the deformation parameter
q very close to 1, then $Q$ is, so to say, " almost dense " in any interval
contained in ${\bf R}^+$,
i.e. $a$ can be approximated quite well by an element of Q. In other terms,
at a macroscopic scale (i.e. for $a\in {\bf R}^+$ such that
$|{ln(a)\over \ln(q)}|\gg 1$) deviations from
the classical scaling property would not be detectable, even
though they would be relevant at microscopic
ones (i.e. for $a\in {\bf R}^+,~|{ln(a)\over \ln(q)}|\sim 1$).
This surprising feature might be considered as a very interesting indication
of the occurrence
of a dishomogeneity of the observable properties of space when the usual
euclidean commutative space is replaced by the corresponding quantum space.

{}~~~Let us consider now eq. (4.29) from the dimensional viewpoint. The series
expansion of the function $f$ makes sense only if $f(x)$ is of the form
$f(x)=g(c_0x)$, where $c_0$ is some constant with dimension of inverse lenght,
$[c_0]=L^{-1}$ (in the case of the harmonic oscillator we can take $c_0=\sqrt
{\omega}$). For the sake of brevity assume that $[g]=1$. Since the volume
form has dimension $L^N$, then $[<f(x)>]=[<g(c_0x)>]=L^N$; this implies
$$
<f(x)>=<g(c_0x)>\propto c_0^{-N}                               \eqno (4.32)
$$
We cannot require property (4.32) to be valid for a scaling
$c_0\rightarrow c'_0:=\alpha c_0$ of the fundamental constant
with an arbitrary $\alpha\in {\bf R}^+$, otherwise
the scaling property of the integral would hold for an arbitrary scaling
parameter as well. The scaling is not incompatible with the ``quantized''
scaling property (4.29) only
if $c'_0\in{\cal I}_{c_0}:=\{c=q^mc_0:~~m\in {\bf Z}\}$. In other words
choosing an integration implies a restriction of the admitted values
for the fundamental constants characterizing the system.

{}~~

{}~~

\centerline{\bf {\large 5}. The Hilbert space of the harmonic oscillator}
\centerline{\bf and the observables $R^i,~P_j,~H_{\omega}$}

{}~~~In this Section we define the pre-Hilbert space of states of the harmonic
oscillator on ${\bf R^N_q}$ and define the observables Hamiltonian, position
and
momentum. We first generate the space through the application of creation
operators to the ground state, then endow it with a scalar product which
mixes the barred and unbarred representation; the latter is designed to make
differential operators (such as the hamiltonian) hermitean in a straightforward
way. The second part is technical and may be skipped by the reader without
serious consequences for the general understanding.
It starts just after formula (5.20) and deals with the
determination of some coefficients which appear in the definition of the
creation/destruction operators.

{}~~~We introduce the pre-Hilbert space ${\cal H}$ of the
$SO_q(N)$-symmetric (isotropic) harmonic oscillator with characteristic
constant $\omega$ in the following way. Let $|0>$ be the ground
state with the energy $E_0$ given in formula (2.36). We introduce a direct
$(\Pi,V)$ and a barred $(\bar\Pi,\bar V)$ representation ($V,\bar V$ were
defined in Sect. 2) by first assuming
$$
\matrix
{& & & & exp_{q^2}[-{\omega q^{-N}(xCx)\over \mu}]\in V & \cr
 & & {\Pi \atop \nearrow} & & & \cr
 &|0>\in {\cal H}& & & & \cr
 & & {\bar\Pi \atop \searrow} & & & \cr
 & & & & exp_{q^{-2}}[-{\omega q^N(xCx)\over {\bar\mu}}]\in \bar V. \cr}
                                                              \eqno (5.1)
$$
Creation and destruction operators $A^{+i},
A^i$ are to be represented respectively by
$$
\matrix
{& & & & b_n(q)(x^i-{q^{2-n}\over \omega}\partial^i)G_q\equiv a_n^{+i} &\cr
 & & {\Pi \atop \nearrow} & & & \cr
 &A^{+i}& & & & \cr
 & & {\bar\Pi \atop \searrow} & & & \cr
 & & & & b_n(q^{-1})(x^i-{q^{n-2}\over \omega}\bar\partial^i)G_{q^{-1}}\equiv
 \bar a_n^{+i}&\cr}
                                                                \eqno (5.2)
$$
when acting on states of level $(n-1)$ (to give states of level $n$), and by
$$
\matrix
{& & & & d_n(q)(x^i+{q^{n+N}\over \omega}\partial^i)G_q\equiv a^i_n &\cr
 & & {\Pi \atop \nearrow} & & & \cr
 &A^i& & & & \cr
 & & {\bar\Pi \atop \searrow} & & & \cr
 & & & & d_n(q^{-1})(x^i+{q^{-n-N}\over \omega}\bar\partial^i)G_{q^{-1}}
\equiv \bar a^i_n &\cr}                                       \eqno (5.3)
$$
when acting on states of level $(n-1)$ (to give states of level $(n-2)$);
the operator $G_q$ was defined in formula (2.21), and the coefficients
$b_n,d_n$ will be fixed below. The space ${\cal H}_n$ of
states of level $n$ will be introduced as linear span of the vectors
$$
\matrix
{& & & & \psi_n^{i_ni_{n-1}...i_1} &\cr
 & & {\Pi \atop \nearrow} & & & \cr
 &|i_n,i_{n-1},...i_1>:=A^{+i_n}A^{+i_{n-1}}...A^{+i_1}|0>& & & & \cr
 & & {\bar\Pi \atop \searrow} & & & \cr
 & & & & \bar \psi_n^{i_ni_{n-1}...i_1} &\cr}.                 \eqno (5.4)
$$
The vector $|i_n,...i_1>$ can be assigned the $SO_q(N,{\bf R})$
transformation law
$$
\phi_l(|i_n,...i_1>)=T^{i_n}_{j_n}...T^{i_1}_{j_1}\otimes |j_n,...j_1>
                                                                 \eqno (5.5)
$$
since both $\psi_n^{i_n...i_1}$ and $\bar\psi_n^{i_n...i_1}$ have
transformation
laws of this kind.
Any $|u_n>\in {\cal H}_n$ is an eigenvector with eigenvalue
$$
E_n=\omega{1\over 2}(q^{{N\over2}-1}+q^{1-{N\over2}})[{N\over 2}+n]_q,
{}~~~~~n\ge 0                                                        \eqno
(5.6)
$$
of the hamiltonian $H_{\omega}$, which is represented by
$$
\matrix
{& & & & h_{\omega}={1\over 2}(-q^N\Delta+\omega^2(xCx)) & \cr
 & & {\Pi \atop \nearrow} & & & \cr
 &H_{\omega}& & & & \cr
 & & {\bar\Pi \atop \searrow} & & & \cr
 & & & & h_{\omega}={1\over 2}(-q^{-N}\bar\Delta+\omega^2(xCx));& \cr}
                                                                   \eqno (5.7)
$$
${\cal H}$ itself is defined as
$$
{\cal H}:=\bigoplus \limits_{n=1}^{\infty}{\cal H}_n.               \eqno (5.8)
$$
By the above construction any vector $|u>\in {\cal H}$ will be represented both
by a vector $\psi_u\in V$ and by a vector $\bar\psi_u\in \bar V$:
$$
\matrix
{& & & & \psi_u & \cr
 & & {\Pi \atop \nearrow} & & & \cr
 &|u> & & & & \cr
 & & {\bar\Pi \atop \searrow} & & & \cr
 & & & & \bar \psi_u. & \cr}                              \eqno (5.9)
$$
With reference to the notation of Sect. 2., we know that any function of the
type $\psi_u=
P_n(x)exp_{q^2}[-{\omega q^{-n-N-2m}xCx \over \mu}]$ belongs to $V$. From the
above construction the corresponding $\bar\psi_u:=\bar \Pi \Pi^{-1}\psi_u\in
\bar V$ will be of the form $\bar\psi_u=
\bar P_n(x)exp_{q^{-2}}[-{\omega q^{+n+N+2m}xCx \over \bar\mu}]$
where the polynomial $\bar P_n(x)$ is obtained from $P_n(x)$ by the following
steps: 1) writing $P_n(x)exp_{q^2}[-{\omega q^{-n-N-2m}xCx \over \mu}]$
as a combinations of the $\psi_m$'s of formula $(5.4)_a$; 2) replacing
$\psi_m$'s by $\bar \psi_m$'s.
If we consider the explicit form of $\psi_m,\bar\psi_m$ involving only the
coordinates (without derivatives) the second step amounts to the substitutions
$q\leftrightarrow q^{-1},~\hat R \leftrightarrow \hat R^{-1}$; in particular
if the $\hat R, \hat R^{-1}$ matrices are written in terms of the projectors
${\cal P}_S,{\cal P}_A,{\cal P}_1$ alone, then we only need to interchange
$q$ with $q^{-1}$.

{}~~~We define the scalar product of two vectors $|v>,|u> \in {\cal H}$ by
the sum of two `` conjugate '' terms:
$$
(u,v):=<\bar\psi_u^*\psi_v> + < \psi^*_u \bar \psi_v >.        \eqno (5.10)
$$
Indeed (~~,~~) is manifestly sesquilinear and (using property (3.4))
$$
(v,u)^*=<\bar\psi_v^*\psi_u>^* + < \psi^*_v \bar \psi_u>^*
=<\psi_u^*\bar\psi_v> + < \bar\psi^*_u \psi_v >=(u,v)         \eqno (5.11)
$$
as required (see relation (3.15)).
Relation (5.11) implies that $(u,u)\in {\bf R}$; its positivity
(i.e. $(u,u)\ge 0$ and $(u,u)=0 \Leftrightarrow u=0$) $\forall q\in {\bf R}^+$
will be proved in Sect. 7.  Here we just note that it must hold at least
in a
($|u>$-dependent) neighbourhood of q=1, as it holds for q=1 and $(u,u)$ is a
continuous function of q.

{}~~~The abstract definition of the hermitean conjugate
$T^{\dag}$ of an operator $T$ is the usual one
$$
(u,Tv)=(T^{\dag}u,v).                                              \eqno (5.12)
$$
We have chosen for the scalar product the (apparently cumbersome) form (5.11)
to make the operator $H_{\omega}$ hermitean. Let us check that this is the
case.
Using the notation introduced in formula (2.27)
$$
\Delta f=f'(x)+ f_j(x,\partial)\partial^j=:\Delta f|+f_j(x,\partial)\partial^j,
{}~~~~~~~~~f\in Fun({\bf R_q^N})                              \eqno (5.13)
$$
and the relation $\bar \Delta=q^{2N}\Delta^*$ it is straightforward to show
that
$$
(\Delta f|)^*g:=f^{'*}g=f^*\cdot(q^{-2N}\bar\Delta g|)-\partial^{j*}f^{*}_jg|.
                                                                    \eqno
(5.14)
$$
Hence
$$
<\psi_u^*\bar h_{\omega}\bar\psi_v>=
<\psi_u^*(\omega^2xCx-q^{-N}\bar\Delta)\bar\psi_v|>          \eqno  (5.15)
$$
and
$$
<(h_{\omega}\psi_u)^*\bar\psi_v>=-q^N<(\Delta \psi_u|)^*\bar\psi_v>+
<\omega^2xCx\psi_u^*\bar \psi_v>
$$
$$
=-q^{-N}<\psi_u^*\bar\Delta \bar\psi_v|>+<\omega^2xCx\psi_u^*\bar \psi_v>
+q^N<\partial^{j*}\psi^{*}_{uj}\bar\psi_v|>;
                                                              \eqno (5.16)
$$
The last term vanishes because of Stoke's theorem (3.7) (in fact
$\partial^{j*}$ are derivatives of $\bar \partial$ type), therefore
$<\psi_u^*\bar h_{\omega}\bar\psi_v>=<(h_{\omega}\psi_u)^*\bar\psi_v>$.
Similarly one proves that
$<\bar\psi_u^*h_{\omega}\psi_v>=<(\bar h_{\omega}\bar\psi_u)^*\psi_v>$.
Hence we find the

{}~

{\bf Proposition 5.1:} the Hamiltonian $H_{\omega}$ is hermitean:
$$
(u,H_{\omega}v)=(H_{\omega}u,v).                                \eqno (5.17)
$$

{}~~~As an immediate consequence of the hermiticity of the hamiltonian,
if $|u>,|v>$ are two
eigenvectors of $H_{\omega}$ with different eigenvalues, then
$$
(u,v)=0.                                                         \eqno (5.18)
$$
Looking back at the previous proof we see that in fact a stronger property
holds:
$$
n\neq m ~~~\Rightarrow~~~~~~~~~~~~~
<\bar\psi_n^*\psi_m>=0,~~~~~<\psi_n^*\bar\psi_m>=0~~~~~\psi_p\in \Psi_p,~~
\bar \psi_p\in \bar\Psi_p                                      \eqno (5.19)
$$
($\Psi_p,\bar \Psi_p$ were defined in formula (2.44))

{}~~~For the evaluation of the scalar products $(~~,~~)$ it is only necessary
to find out integrals of the type $<(xCx)^k f(xCx)>$ with
$$
f=exp_{q^{-2}}[-{\omega q^{N+k+2m}xCx \over \bar\mu}]
exp_{q^2}[-{\omega q^{-N-k-2m}xCx \over\mu}],                   \eqno (5.20)
$$
since their tensor structure is already determined by the general knowledge
of the tensors $S^{i_1...i_{2n}}$ of Section 4.; these integrals will
be determined in Appendix B.

{}~~~We still have to fix the coefficients $b_n,d_n$ to complete the
definitions
(5.2),(5.3) of $A^i,A^{+i}$. We determine them imposing two requirements:
1) that creation/destruction operators are hermitean conjugate of each-other,
equation (5.23); 2) that the position operators do not contain derivatives,
equation (5.34). The final result is shown in equations (5.36),(5.38).
The reader not interested in these computations might simply give a glance to
these formulae. Before starting, we mention two
relations which we will use in doing this job:
$$
<(\bar a_{n+1}^{+i}\bar a_n^{+j}\bar\psi)^*\psi>=q^{-1-N}
{b_{n+1}(q^{-1})b_n(q^{-1})\over d_n(q)d_{n-1}(q)}
<\bar \psi^*a_{n-1}^{j'}a_n^{i'}\psi>C_{j'j}C_{i'i}      \eqno (5.21)
$$
$$
<(a_{n+1}^{+i}a_n^{+j}\psi)^*\bar\psi>=
q^{1+N}{b_{n+1}(q)b_n(q) \over d_n(q^{-1})d_{n-1}(q^{-1})}
<\psi^*\bar a_{n-1}^{j'}\bar a_n^{i'}\bar\psi>
C_{j'j}C_{i'i};                                                  \eqno (5.22)
$$
they can be derived using Stoke's theorem (3.7) and the scaling property
(4.31).

{}~~~First we require that
$$
(A^{+i})^{\dag}=A^lC_{li}~~~~~~~~~~(A^i)^{\dag}=A^{+l}C_{li};   \eqno (5.23)
$$
using the orthogonality relations (5.18) this requirement reduces to
$$
(A^{+i}u_n,u_{n+1})=(u_n,A^lu_{n+1})C_{li}~~~~~~~~
$$
$$
(A^iu_{n+1},u_n)=(u_{n+1},A^{+l}u_n)C_{li},~~~~~~~~\forall n\ge 0,~~~~
\forall u_n\in {\cal H}_n,~~~u_{n+1}\in {\cal H}_{n+1}.           \eqno (5.24)
$$
Condition (5.24) is equivalent to
$$
\cases {(A^{+i}A^{+j}u_m,u_{m+2})=(u_m,A^kA^lu_{m+2})C_{kj}C_{li}~~~~\forall
m\ge 0,\cr
(A^iA^ju_{m+2},u_m)=(u_{m+2},A^{+k}A^{+l}u_m)C_{kj}C_{li},~~~~~~~~
\forall u_m\in {\cal H}_m,u_{m+2}\in {\cal H}_{m+2}    \cr}
$$
$$
(A^{+i}u_0,u_1)=(u_0,A^lu_1)C_{li}~~~~~~~~
(A^iu_1,u_0)=(u_1,A^{+l}u_0)C_{li}                              \eqno (5.25)
$$
The implication $(5.24)\Rightarrow (5.25)$ is trivial. To prove the converse
one
we just need to express the vectors $|u_n>$ as combinations of the vectors
(5.4). Writing down conditions $(5.25)_a$ explicitly in terms of $b_n,d_n$ and
using relations (5.21),(5.22), we can check that these conditions are
satisfied if
$$
{b_{m+2}(q^{-1})b_{m+1}(q^{-1})\over q^{N+1}d_{m+1}(q)d_m(q)}
<\bar \psi_{u_m}^*a_m^{j'}a_{m+1}^{i'}\psi_{u_{m+2}}>=<\bar \psi_{u_m}^*
a_{m+2}^{j'}a_{m+3}^{i'}\psi_{u_{m+2}}>~~~~~~\forall q\in{\bf R}^+, \eqno
(5.26)
$$
and similarly for the conjugate term. According to
formula (5.19) we can replace in the RHS of (5.26) the function
$a_m^{j'}a_{m+1}^{i'}\psi_{u_{m+2}}$ by the one $P_{\Psi_m}
[a_m^{j'}a_{m+1}^{i'}\psi_{u_{m+2}}]$ ($P_{\Psi_m}$ denotes the projector
onto the space $\Psi_m$). Using formula (B.9) we can check that
$$
P_{\Psi_m}[a_m^{j'}a_{m+1}^{i'}\psi_{u_{m+2}}]=
$$
$$
={q^{m+1+N}+q^{-m-1}\over q^{m+3+N}+q^{-1-m}}{q^{m+N}+q^{-m}\over
q^{m+2+N}+q^{-m}}{d_{m+1}(q)d_m(q)\over d_{m+3}(q)d_{m+2}(q)}
a_{m+2}^{j'}a_{m+3}^{i'}\psi_{u_{m+2}}                            \eqno (5.27)
$$
Collecting this information we find that (5.26) and hence $(5.25)_a$
are satisfied if
$$
{b_{m+2}(q^{-1})\over d_{m+3}(q)}=q^{N+1}{1+q^{2m+4+N}\over 1+q^{2m+N}}
{d_{m+2}(q)\over b_{m+1}(q^{-1})}~~~~~~~~~\forall m\ge 0,~~~
\forall q\in{\bf R}^+.                                             \eqno (5.28)
$$
As for conditions $(5.25)_b$, an explicit computation shows that they are
satisfied if
$$
d_2(q)=b_1(q^{-1})q^{-N-{1\over 2}}{1+q^N\over 1+q^2}\phi(q)
{}~~~~~~~~~~~~~\forall q\in{\bf R}^+,                                \eqno
(5.29)
$$
where the constant $\phi(q)$ is defined in formula $(B.8)$. Solving the
recursive equation (5.28) by taking relation (5.29) as the initial condition,
we find
$$
{d_{m+2}(q)\over b_{m+1}(q^{-1})}=\left [{(1+q^2)q^{N\over 2}\over \phi
(1+q^{2+N})} \right ]^{(-1)^{m+1}} q^{-N-1\over 2}{1+q^{2m+N}\over
1+q^{2m+2+N}}~~~~~~~~~\forall q\in{\bf R}^+,~~~~m\ge 0.           \eqno (5.30)
$$
Summing up, relation (5.30) guarantees that equation (5.23) is satisfied.

{}~~~As a direct consequence of equation (5.23), if $f_l\in {\bf C}$ are
numbers such that
$$
C_{ml}f^*_l=f_m,                                                  \eqno (5.31)
$$
then the operators $f_l(A^l+A^{+l}),{f^l\over i}(A^l-A^{+l})$ are hermitean
operators. There exist $N$ independent solutions $f^i_l,~i=1,2,...,N$ of
equations (5.31). For instance if $N=3$ we can take
$$
\Vert f^i_l \Vert={1\over 2}
\left \Vert\matrix
{& 1 & 0 & q^{1\over 2} &\cr
 & 0 & 1 & 0            &\cr
 &{1\over i} & 0 & iq^{1\over 2} &\cr}\right \Vert.        \eqno (5.32)
$$
In general, ,given $N$ solutions $f^i_l$ of $(5.31)$ we define
$$
R^i:={1\over\sqrt{\omega}}f^i_l(A^l+A^{+l})~~~~~~~~~~
P^i:={1\over i\sqrt{\omega}}f^i_l(A^l-A^{+l}).                    \eqno (5.33)
$$
Additional conditions on the coefficients $b_n,d_n$ arise if we impose
the requirement that the operators $\Pi(R^i),\bar \Pi(R^i)$ (acting
respectively
in $V,\bar V$) contain no derivative:
$$
\matrix
{& & & & g_n(q)f^i_lx^l G_q~on~\Psi_n &\cr
 & & {\Pi \atop \nearrow} & & & \cr
 & R^i & & & & \cr
 & & {\bar\Pi \atop \searrow} & & & \cr
 & & & & g_n(q^{-1})f^i_lx^l G_q~on~\bar\Psi_n &\cr}.          \eqno (5.34)
$$
Equation (5.34) and the definitions (5.2),(5.3) of $A^l,A^{+l}$ imply
$$
d_{m+1}(q)=q^{-2m-N}b_{m+1}(q),~~~~~~~~~g_m(q)={1+q^{-2m-N}\over \sqrt{\omega}
}b_{m+1}(q).                                                       \eqno (5.35)
$$
Relations (5.30), (5.35) determine all the coefficients $b_n,d_n$, $n\ge 1$,
in terms of $b_1$. The solutions are
$$
d_n(q)=q^{2-2n-N}b_n(q),~~~~~~~b_n(q)={(1+q^N)q^{2n-2}\over 1+q^{2n-2+N}}\cdot
\cases {b_1(q)~~if~n~is~odd \cr
       \phi {1+q^{N+2}\over 1+q^2}q^{-1\over 2}b_1(q^{-1})~~if~n~is~even  \cr}.
                                                                \eqno (5.36)
$$
Consequently
$$
g_n=\cases{{1+q^{-N}\over \sqrt{\omega}}b_1(q)\equiv g_+~~~if~n~is~even  \cr
        {(1+q^N)(1+q^{-2-N})\over (1+q^2)\sqrt{\omega}}q^{3\over 2}\phi
        b_1(q^{-1})\equiv g_-~~~if~n~is~odd \cr}               \eqno (5.37)
$$
We are still free to fix $b_1$. In the sequel we will take
$$
b_1(q)={\sqrt{2\omega}\over 1+ q^{-N}}~~~~~~\Rightarrow~~~~~~~g_+=\sqrt{2},
{}~~~g_-=\phi q^{3\over 2}{1+q^{-2-N}\over 1+q^2}\sqrt{2}.
\eqno(5.38)
$$
The above choice is such that $b_n(1)=\sqrt{\omega\over 2}=d_n(1)$, so that
in the classical limit $q=1$ $A^{+i},A^i$ give the classical creation/
annihilation operators. Let
$$
{\cal H}^+:=\sum \limits_{h=0}^{\infty}{\cal H}_{2h}~~~~~~~~~
{\cal H}^-:=\sum \limits_{h=0}^{\infty}{\cal H}_{2h+1}              \eqno
(5.39)
$$
We can summarize formulae (5.34),(5.37) by
$$
\matrix
{& & & & g_{\pm}(q)f^i_lx^l G_q \psi_u & & & & & & \cr
 & & {\Pi \atop \nearrow} & & & & & & & & \cr
 & R^i|u>& & & & & & & & if~|u>\in{\cal H}_{\pm} & \cr
 & & {\bar\Pi \atop \searrow} & & & & & & & & \cr
 & & & & g_{\pm}(q^{-1})f^i_lx^l G_{q^{-1}} \bar\psi_u & & & & & & \cr},
                                                                    \eqno
(5.40)
$$
where $\psi_u=\Pi(|u>)\in V_{\pm}$, $\bar\psi_u=\bar\Pi(|u>)\in \bar V_{\pm}$.
It is natural to call the observables $R^i$ the position operators of the
system, since they reduce to the ordinary position operators when $q=1$.
Because of our requirement (5.34) $\Pi(R^i)G_{q^{-1}}$,
$\bar\Pi(R^i)G_q$ act as pure
multiplication by a combination of coordinates $x^i$. The classical commutation
relations $[R^i,R^j]=0$ are replaced by the new ones
$$
\tilde {\cal P}_{A~hk}^{~~ij}R^hR^k=0,~~~~~~~~~~~~~
\tilde{\cal P}_A:=(f\otimes f){\cal P}_A (f^{-1}\otimes f^{-1})     \eqno
(5.41)
$$
Up to a factor there exixts only one quadratic function of the $R^i$'s which
is a scalar and an observable, $R^2:={1\over 2}R^i(f^{-1T}Cf^{-1})_{ij}R^j$
(notice that the matrix $(f^{-1T}Cf^{-1})$ is hermitean), therefore we will
call it the square lenght.
Since the action of $R^i$ flips the parity of a vector, $R^2$
is represented in the same way on all of ${\cal H}$:
$$
\matrix
{& & & & \phi{q^{1+{N\over 2}}+q^{-1-{N\over 2}}\over q+q^{-1}}
          q^{-1-{N\over 2}}xCx G_{q^2} &\cr
 & & {\Pi \atop \nearrow} & & & \cr
 & R^2 & & & & \cr
 & & {\bar\Pi \atop \searrow} & & & \cr
 & & & & \phi{q^{1+{N\over 2}}+q^{-1-{N\over 2}}\over q+q^{-1}}
          q^{1+{N\over 2}}xCx  G_{q^{-2}}&\cr}.          \eqno (5.42)
$$
According to the definition (5.33) the observable $P^i$ will be represented
by
$$
\matrix
{& & & & {1\over i\sqrt{\omega}}b_{n+1}(q)f^i_l[(q^{-2n-N}-1)x^l+2{q^{1-n}\over
 \omega}\partial^l] G_q \psi_{u_n} & & \cr
 & & {\Pi \atop \nearrow} & & & & \cr
 & P^i|u_n>& & & & if~|u_n>\in{\cal H}_n & \cr
 & & {\bar\Pi \atop \searrow} & & & & \cr
 & & & & {1\over i\sqrt{\omega}}b_{n+1}(q^{-1})f^i_l[(q^{2n+N}-1)x^l+2{q^{n-1}
\over \omega}\bar\partial^l] G_{q^{-1}}\bar \psi_{u_n}. & & \cr}
                                                                    \eqno
(5.43)
$$
They will be called momentum operators, since they reduce to the ordinary
momentum operators in ${\bf R}^N$ when $q=1$. Contrary to the classical case,
from formula (5.43) we recognize that $\Pi(P^i)G_{q^{-1}},\bar \Pi(P^i)G_q$
are not pure derivatives. A straightforward computation shows that the
classical commutation relations $[P^i,P^j]=0$ are replaced by the new ones
$$
\tilde {\cal P}_{A~hk}^{~~ij}P^hP^k=0.                            \eqno (5.44)
$$
The reader might ask why we have not defined the position/momentum
operators so that the square lenght and the square momentum be represented
in $V$ (resp. in $\bar V$) by $xCx, q^N\Delta$ (resp. $xCx,q^{-N}\bar \Delta$).
The reason is that
the operators $xCx, q^N\Delta$ (resp. $xCx,q^{-N}\bar \Delta$) do not
map all of $V$ (resp. $\bar V$) into itself (for instance, if $\psi_0$ denotes
the ground state in $V$, then it is easy to check that $(xCx)\psi_0\notin V$).
Therefore the hamiltonian (5.7) cannot be written as a combination of the
square lenght and square momentum. This difficulty seems difficult to overcome
by modifying the hamiltonian and therefore its eigenfunctions and $V$ itself.
For instance, looking
for the eigenfunctions of a hamiltonian of the type $q^{1+{N\over 2}}
(\omega^2 xCx-q^{-2-N}\Delta)G_{q^2}$, which is formally hermitean,
one can find that they are functions belonging to the space
$F:=\{f=P(x)exp_q^2[-{\omega q^{1-{N\over 2}}xCx\over \mu}]\}$: here,
contrary to the
eigenfunctions $\psi_n$ of $h_{\omega}$, the exponent in the gaussian is
independent of the degree $n$ of the polynomial $P(x)$. One immediately
realizes that the operator $q^{1+{N\over 2}}xCxG_{q^2}$ maps $F$ out of itself,
i.e. it is not a well defined operator in $F$; $xCx$, instead, would be.
Therefore it would be natural to consider the latter as the representative in
$F$ of the square lenght.
Similarly, $q^{-1-{N\over 2}}\Delta G_{q^2}$ is not a well defined operator
in $F$, whereas $q^{-N}\Delta$ is. Thus again we would
see that this new hamiltonian cannot be written as a combination of the square
lenght and of the square momentum.

{}~~

{}~~

\centerline{\bf {\large 6}. The observables $L^2,L_m$}

{}~~

{}~~~We look for some other hermitean operators such that they
commute
with the hamiltonian $H_{\omega}$ and with each other. To this end in this
section we search the analog of the angular momentum. We will mainly give
explicit definitions and relations in the unbarred representation; the barred
ones can be found performing the usual substitutions.

{}~~~As a primary requirement, the components of the angular momentum should
commute with any scalar function
of the coordinates and of the momenta. In the classical case they are
antisymmetrized products of coordinates and derivatives of the type ${1\over i}
(y^i\partial^{y^j}-y^j\partial^{y^i})$ or their combinations. Therefore we
first look at the commutation relations of the operators
${\cal L}^{ij}:={\cal P}^{~~ij}_{A~hk}x^h\partial^k=
-q^{-2}{\cal P}^{~~ij}_{A~hk}\partial^hx^k$  with $xCx,\Delta$. Using
formula (2.31) we find
$$
{\cal L}^{ij}xCx=q^2xCx{\cal L}^{ij},~~~~~~~
{\cal L}^{ij}\Delta=q^{-2}\Delta{\cal L}^{ij}.                       \eqno
(6.1)
$$
It immediately follows that
$$
[G_{q^2}{\cal L}^{ij},xCx]=0=[G_{q^2}{\cal L}^{ij},\Delta]      \eqno (6.2)
$$
Next, it is easy to show that $G_{q^2}{\cal L}^{ij}$
commutes with any scalar polynomial $I$ (in particular $h_{\omega}$)
obtained combining $x^i$'s and $\partial^i$'s:
$$
[I(x,\partial),G_{q^2}{\cal L}^{ij}]=0.                          \eqno (6.3)
$$
Actually any such
polynomial can be written as a polynomial in $xCx$, $\Delta$ alone
(see Appendix C).
For this reason, we propose the operators $G_{q^2}{\cal L}^{ij}$'s
(and their barred partners) as candidates to the role of angular momentum
components.

{}~~~By squaring the ${\cal L}^{ij}$  we obtain the
scalar operator ${\cal L}^2$:
$$
{\cal L}^2:={\cal L}^{ij}{\cal L}_{ji}=x^h\partial^k{\cal P}_{A~hk}
^{~~ij}x_j\partial_i                                        \eqno (6.4)
$$
To obtain the last expressions in (6.4) we have used the property
(2.8) (where we take $f(\hat R)={\cal P}_A$), and ${\cal P}_A^2={\cal P}_A$.
Of course $G_{q^4}{\cal L}^2$ commutes
with any scalar function $I(x,\partial)$ and in particular with $h_{\omega}$.

{}~~~We want to find out eigenvalues and eigenfunctions of $G_{q^4}{\cal L}^2$.
{}From the above property it is clear that if $P(x)$ is an eigenvector
of $G_{q^4}{\cal L}^2$, then for any function $p=p(xCx)$ $g:=P(x)p(xCx)$ is
an eigenvector of $G_{q^4}{\cal L}^2$ with the same eigenvalue.
A little thinking will convince the reader that, just as in the classical
case, after factorizing a possible function $p(xCx)$ the eigenvectors
$P(x)$ of $G_{q^4}{\cal L}^2$
$$
G_{q^4}{\cal L}^2 P(x)|= c P(x)                               \eqno (6.5)
$$
can be written as homogeneous polynomials:
$$
P(x)=p(xCx)A_{i_1i_2...i_n}x^{i_1}x^{i_2}...x^{i_n}         \eqno (6.6)
$$
Actually $G_{q^4}{\cal L}^2$ is homogeneous
in both $x$, $\partial$ with the same degree (two),
hence it transforms any homogeneous polynomial of degree $n$ into another
one of the same degree.
Now we specify the form of the $A_{i_1i_2...i_n}$ coefficients.
We are going to prove that just as in the classical case
and up to factors $p(xCx)$ the set of homogeneous polynomials
$$
P_S^{l_1l_2...l_n}:={\cal P}^{~~~~l_1l_2...l_n}_{n,S~i_1i_2...i_n}
x^{i_1}x^{i_2}...x^{i_n},~~~l_i=1,...,N                       \eqno (6.7)
$$
is a complete set of eigenvectors of degree $n$ of both
$G_{q^4}{\cal L}^2$ and $G_{q^{-4}}\bar{\cal L}^2$. Here ${\cal P}_{n,S}$
is the q-deformed symmetric projector acting on $\bigotimes^n {\bf C}$
(in particular ${\cal P}_{2,S}\equiv {\cal P}_S$), whose properties will be
briefly discussed in Appendix D.
The main property of these projectors is that
$$
{\cal P}_{n,S}{\cal P}_{A~i,(i+1)}=0={\cal P}_{n,S}{\cal P}_{1~i,(i+1)},
{}~~~~~~~~~1 \le i \le n-1                                      \eqno (6.8)
$$
where, for any matrix $F$ defined on $C\otimes C$ we denote by
$F_{i,i+1}~~(1\le i\le n-1)$ the matrix defined by
$F_{i,i+1}:= {\bf 1}\otimes...{\bf 1}\otimes F \otimes {\bf 1}
\otimes...\otimes {\bf 1}$ ($F$ at the $i^{th}$ and $(i+1)^{th}$ place).
Since all the projectors are symmetric, the above properties hold also if
we multiply ${\cal P}_{n,S}$ by ${\cal P}_{A~i,(i+1)},~{\cal P}_{1~i,(i+1)}$
from the left. Relations (6.8) imply
$$
{\cal P}_{n,S}{\cal P}_{S~i,(i+1)}={\cal P}_{n,S}
{}~~~~~~~~~1 \le i \le n-1.                                \eqno (6.9)
$$

{}~~~Let us consider the space $M_n$ of homogeneous polynomials
of degree $n$ (see formula 2.51) and its two projections
$$
M_{n}^S={\cal P}_{n,S} M_n~~~~~~~~~~~~~~~M_n^1=({\cal P}_1\otimes
{\bf 1}_{n-2})M_n.                                            \eqno (6.10)
$$
{}~

{\bf Proposition 6.1:} $M_n$ can be decomposed into the direct sum
$$
M_n=\bigoplus \limits_{0\le m\le {n\over 2}} M^S_{n,n-2m}~~~~~~~~
M^S_{n,n-2m}:=(xCx)^m M_{n-2m}^S.                            \eqno (6.11)
$$
In other words
$$
{\bf 1}_{M_n}:=\bigoplus \limits_{0\le m\le {n\over 2}}
({\cal P}_1\otimes...\otimes {\cal P}_1\otimes {\cal P}_{n-2m,S}) \eqno (6.12)
$$
is the identity operator on $M_n$.

$ Proof:$ Let us consider $M_k$. Because of formula (2.9),(6.8)
$$
M_k=M_{k}^S\oplus M_k^1=[{\cal P}_{k,S}\oplus ({\cal P}_1\otimes
{\bf 1}_{k-2})]M_k.                                            \eqno (6.13)
$$
By repeated application of formula (6.13) for $k=n,n-2,...$ we arrive at the
claim $\diamondsuit$.

{}~

We can evaluate the dimensions of the spaces $M_S^k$ in a
straightforward way, since we know the dimension of $M_l$ as a function of $l$
(see formula $(2.50)_b$) and $M_k^1$ is generated by
$\{x^{i_1}...x^{i_{k-2}}x^{j_{k-1}}x^{j_k}
{\cal P}_{1~j_{k-1}j_k}^{~~i_{k-1}i_k}\}$, i.e. by
$\{x^{i_1}...x^{i_{k-2}}(xCx)\}$:
$$
dim(M_k^1)=dim(M_{k-2})={N+k-3 \choose N-1}
$$
$$
dim({\cal P}_k^S):=dim(M_k^S)=dim(M_k)-dim(M_k^1)
=dim(M_k)-dim(M_{k-2})                                        \eqno (6.14)
$$

{}~

{\bf Proposition 6.2}: $M_n^S$ are eigenspaces of $2q^2G_{q^4}{\cal L}^2$,
$2q^{-2}G_{q^{-4}}\bar{\cal L}^2$ with the same eigenvalue
$$
l_n^2={2\over q+q^{-1}}{q^{{N\over 2}-2}+q^{2-{N\over 2}} \over
  q^{{N\over 2}-2}+q^{1-{N\over 2}}} [n]_q[N+n-2]_q.          \eqno (6.15)
$$
{}~

Note: we have included in the definition of these operator a factor 2 so that
for $N=3$ and
q=1 the eigenvalues reduce to the classical ones $n(n+1)$ of the classical
square angular momentum in three dimensions. Moreover, as the energies
$E_n$, the eigenvalues
(6.15) are invariant under the transformation $q\rightarrow q^{-1}$.

$Proof:$ To reach the goal we first transform ${\cal L}^2$ into a
more suitable form, which explicitly shows its scalar character.
By quite a lenghty calculation one can show that
$$
{\cal L}^2=\alpha_N(q)x^i\partial_i+\beta_N(q)x^ix^j\partial_j\partial_i
+\gamma_N(q)xCx\Delta                                    \eqno (6.16)
$$
where
$$
\alpha_N(q):={(q^{2-{N\over 2}}+q^{{N\over 2}-2})\over (q^{1-{N\over 2}}+
q^{{N\over 2}-1})}  {(q^{1-N}-q^{N-1})\over (q^{-2}-q^2)},
$$
$$
\beta_N(q):={q^3+q^{N-1} \over \mu (q+q^{-1})},~~~~~~~~~
\gamma_N(q):=-{(q^{5-N}+q)(1+q^{-N}) \over \mu^2 (q+q^{-1})}.    \eqno (6.17)
$$
Now notice that property (6.9) together with the relations
$$
{\cal P}^{~~ij}_{S~hk}x^h\partial^k={\cal P}^{~~ij}_{S~hk}\partial^h x^k,~~~~
                                                             \eqno (6.18)
$$
implies
$$
{\cal
P}_{n,S~b_1b_2...b_n}^{~~~~a_1a_2...a_n}x^{b_1}...x^{b_{i-1}}\partial^{b_i}
x^{b_{i+1}}...x^{b_n}|=0.                                     \eqno (6.19)
$$
Similarly, upon use of formula (2.31)
$$
{\cal P}_{n,S~b_1b_2...b_n}^{~~~~a_1a_2...a_n}x^{b_1}...x^{b_{i-1}}
\Delta x^{b_i}...x^{b_n}|=0                               \eqno (6.20)
$$
This means that when applying ${\cal L}^2$
to ${\cal P}^{~~~~l_1l_2...l_n}_{n,S~i_1i_2...i_n}x^{i_1}x^{i_2}...x^{i_n}$
we can forget all the terms (which we will denote by dots) containing powers
of $\Delta$ or where the index $b_i$ of a derivative
$\partial^{b_i}$ is contracted with an index
of ${\cal P}_{n,S}$. The term with coefficient $\gamma_N$ in the RHS of (6.16)
can be ignored, whereas
$$
(x^a\partial_a)x^{b_i}=x^{b_i}+q^2x^{b_i}(x^a\partial_a)+...
$$
$$
(x^ax^b\partial_b\partial_a)x^{b_i}=(1+q^2)x^{b_i}(x^a\partial_a)+q^4 x^{b_i}
(x^ax^b\partial_b\partial_a)+...                                 \eqno (6.21)
$$
By a somewhat lenghty calculation we find
$$
G_{q^4}{\cal L}^2P_S^{l_1l_2...l_n}=c_nP_S^{l_1l_2...l_n}
                                                                 \eqno (6.22)
$$
where
$$
c_n:={q^{-2}\over q+q^{-1}}
{q^{{N\over 2}-2}+q^{2-{N\over 2}} \over
  q^{{N\over 2}-1}+q^{1-{N\over 2}}} [n]_q[N+n-2]_q;             \eqno (6.23)
$$
$P_S^{l_1...l_n}(x)$ are therefore eigenvectors of $G_{q^4}{\cal L}^2$, with
eigenvalues depending only on $n$. In the derivation
we have used the formula $(2.28)_b$ and the relation
$$
\sum \limits_{k=1}^n q^{2k} k_{q^2}=q^2 {n_{q^2}(n+1)_{q^2} \over 2_{q^2}},
                                                          \eqno (6.24)
$$
(the latter can be easily proved iteratively).
In the same way we can show that the eigenvalues $\bar c_n$ of $G_{q^{-4}}
\bar {\cal L}^2$ are given by $\bar c_n=q^4c_n$.
We see that $M^S_n$ is eigenspace of both $2q^2G_{q^4}{\cal L}^2$ and
 $2q^{-2}G_{q^{-4}}\bar{\cal L}^2$ with the same eigenvalue $l^2_n$ given
in formula (6.15) $\diamondsuit$.

{}~

{}~~~As a direct consequence of propositions 6.1,6.2 we find the

{}~

{\bf Corollary 6.3:} relation (6.11) provides the decomposition of $M_n$ into
the direct sum of eigenspaces $M^S_{n,n-2m}$ of
$2q^2G_{q^4}{\cal L}^2$ and $2q^{-2}G_{q^{-4}}\bar{\cal L}^2$ with
eigenvalues $l^2_{n-2m}$, $m=0,1,...[{n\over 2}]$.

{}~

{}~~~Since $[h_{\omega},G_{q^4}{\cal L}^2]=0$, it is possible to find
eigenvectors of $h_{\omega},~G_{q^4}{\cal L}^2$  at the same time. Using
again property (6.9) it is quite easy to realize that
$$
[({\cal P}_1\otimes...\otimes {\cal P}_1\otimes {\cal P}_{n-2m,S})\psi_n]
^{l_1...l_n}\propto
$$
$$
\propto P^{l_{2m+1}...l_n}_S p_{n,m}(xCx)
exp_{q^2}[-{\omega q^{-n-N}xCx\over \mu}],                \eqno (6.25)
$$
(with suitable polynomials $p_{n,m}$, $ 0\le m\le {n\over 2}$), hence these
functions are eigenvectors of $2q^2G_{q^4}{\cal L}^2$ with eigenvalue
$l^2_{n-2m}$. The same holds for the analogous combinations of $\bar \psi$'s.
Using the property (2.49) (${\cal P}_{A~i,i+1}\psi_n=0$) we conclude that
${\bf 1}_{M_n}$ is the identity operator in $\Psi_n,\bar \Psi_n$ as well.
Therefore
$$
\Psi_n=\bigoplus \limits_{0\le m\le {n\over 2}}\Psi_{n,n-2m}~~~~~~~~(resp.
{}~~~\bar\Psi_n=\bigoplus \limits_{0\le m\le {n\over 2}}\bar\Psi_{n,n-2m}),
                                                            \eqno (6.26)
$$
where
$$
\Psi_{n,n-2m}:=({\cal P}_1\otimes...\otimes {\cal P}_1\otimes {\cal
P}_{n-2m,S})
\Psi_n~~~~~~ (resp.~~\bar\Psi_{n,n-2m}:=({\cal P}_1\otimes...\otimes {\cal P}_1
\otimes {\cal P}_{n-2m,S})\Psi_n ),                         \eqno (6.27)
$$
is the eigenspace of $h_{\omega},2q^2G_{q^4}{\cal L}^2$ (resp. of
$\bar h_{\omega},2q^{-2}G_{q^{-4}}\bar{\cal L}^2$) with eigenvalues
$E_n,l^2_{n-2m}$.

{}~~~The above discussion shows that we are in the right condition to define a
square angular momentum operator $L^2$ in ${\cal H}$. We set
$$
\matrix
{& & & & 2q^2G_{q^4}{\cal L}^2  &                      \cr
 & & {\Pi \atop \nearrow} & & &                          \cr
 &L^2 & & & &                                            \cr
 & & {\bar\Pi \atop \searrow} & & &                      \cr
 & & & & 2q^{-2}G_{q^{-4}}\bar{\cal L}^2 &.            \cr}     \eqno (6.28)
$$
We introduce the subspaces ${\cal H}_{n,n-2m}\subset {\cal H}$ by
$$
\matrix
{& & & & \Psi_{n,n-2m} &                      \cr
 & & {\Pi \atop \nearrow} & & &                          \cr
 &{\cal H}_{n,n-2m}& & & &                                            \cr
 & & {\bar\Pi \atop \searrow} & & &                      \cr
 & & & & \bar\Psi_{n,n-2m} &.            \cr}     \eqno (6.29)
$$
We summarize the preceding results in the

{}~

{\bf Proposition 6.4:} ${\cal H}_{n,n-2m}$ ($n\ge 0$, $0\le m\le {n\over 2}$)
is an eigenspace of the operators $H_{\omega},L^2$ defined by (5.7),(6.28)
with eigenvalues $E_n$, $l^2_{n-2m}$ (see (5.6),(6.15)) respectively. Moreover
$$
{\cal H}=\bigoplus\limits_{n=0}^{\infty}\bigoplus \limits_{0\le m \le
{n\over 2}}{\cal H}_{n,n-2m}.                                    \eqno (6.30)
$$

{}~

{}~~~It remains to show that $L^2$ is hermitean. The proof is similar to the
one
we gave for $H_{\omega}$; one uses Stoke's theorem (3.7), the scaling
property (4.31) of $<~~>$ and the relation
$$
{\cal L}^{2*}=q^{-2N-4}\bar {\cal L}^2.                         \eqno (6.31)
$$
The latter can be drawn using relation (2.24), and formula (2.8).
Explicitly:
$$
(L^2u,v)=(u,L^2v).                                             \eqno (6.32)
$$
The direct consequence of formulae (5.23),(6.32)
is that ${\cal H}_{n,m}$ are orthogonal subspaces of ${\cal H}$, i.e.
$$
u\in {\cal H}_{n,k},~~v\in {\cal H}_{n',k'}~~and~~(n,k)\neq(n',k')
\Rightarrow (u,v)=0.                                             \eqno (6.33)
$$
The previous proof actually implies a stronger property:
$$
<\bar\psi_{n,k}^* \psi_{n',k'}>=0=<\psi_{n,k}^* \bar\psi_{n',k'}>
{}~~~~~~~~if~~(n,k)\neq (n',k'),                                \eqno (6.34)
$$
where $\psi_{p,h}\in\Psi_{p,h},~\bar\psi_{p,h}\in\bar\Psi_{p,h}$.

{}~~~Now we show how to construct the observables `` angular momentum
components ''. As already noted at the beginning of this section,
$G_{q^2}{\cal L}^{ij}$ (respectively
$G_{q^{-2}}\bar{\cal L}^{ij}$) commutes with any scalar function
$I(x,\partial)$
(respectively $I(x,\bar\partial)$), in particular with $h_{\omega},G_{q^4}{\cal
L}^2$ (respectively with $h_{\omega},G_{q^{-4}}\bar{\cal L}^2$). We therefore
look for combinations $G_{q^2}{\cal L}_m:=m_{ij}G_{q^2}{\cal L}^ij$,
$G_{q^{-2}}
\bar {\cal L}_{\bar m}:=\bar m_{ij}G_{q^{-2}}\bar
{\cal L}^ij$ which can be considered as the representatives in $V,\bar V$ of
one and the same operator in ${\cal H}$. To this end they should have the same
eigenvalues and there should exist an isomorphism between the corresponding
eigenfunctions.

{}~~~Consider a function $\chi(x)$ which is an eigenvector of $G_{q^2}{\cal
L}_m$:
$$
G_{q^2}{\cal L}_m\chi|=\lambda\chi.                              \eqno (6.35)
$$
Up to a factor $f(xCx)$ $\chi$ must be a homogeneous polynomial of the type
$$
\chi_D:=D_{l_1...l_k}
P_S^{l_1l_2...l_k},~~~~~~~k\ge 1,~~~~D_{l_1...l_k}\in {\bf C},     \eqno (6.36)
$$
since $[g_{q^4}{\cal L}^2,G_{q^2}{\cal L}_m]=0$. Hence equation (6.35)
explicitly reads
$$
G_{q^2}m_{ij}D_{l_1...l_k}{\cal P}^{~~~~l_1l_2...l_k}_{k,S~i_1i_2...i_k}
{\cal P}^{~~ij}_{A~hk}x^h\partial^k x^{i_1}x^{i_2}...x^{i_k}|=\lambda\chi_D.
                                                                   \eqno (6.37)
$$
Using relations (6.8),(6.9) one can easily show that the RHS can be rewritten
in the following way:
$$
G_{q^2}m_{ij}D_{l_1...l_k}{\cal P}^{~~~~l_1l_2...l_k}_{k,S~i_1i_2...i_k}
{\cal P}^{~~ij}_{A~hk}x^hk_{q^2}C^{kj_1}x^{i_2}...x^{i_k}=\lambda\chi_D.
                                                                  \eqno (6.38)
$$
On the other hand, using exactly the same relations and arguments one can
show that $G_{q^{-2}}\bar {\cal L}_m\chi_D|=q^2G_{q^2}{\cal L}_m\chi_D|$,
hence we conclude that $\chi_D$ is an eigenvector of both
$q^{-1}G_{q^{-2}}\bar {\cal L}_m$ and $qG_{q^2}{\cal L}_m$ with the same
eigenvalue. We are led to guess that we can introduce a well defined operator
$L_m$ on ${\cal H}$ by the definition
$$
\matrix
{& & & & qG_{q^2}{\cal L}_m  &                      \cr
 & & {\Pi \atop \nearrow} & & &                          \cr
 &L_m & & & &                                            \cr
 & & {\bar\Pi \atop \searrow} & & &                      \cr
 & & & & q^{-1}G_{q^{-2}}\bar{\cal L}_m &.            \cr}     \eqno (6.39)
$$
Using the same arguments employed when proving the hermiticity of $L^2$ one can
formally show that $L^m$ is hermitean provided the
coefficients $m_{ij}$'s satisfy the condition
$$
[(C\otimes C)\cdot m^*]_{ij}=m_{ji}.                           \eqno (6.40)
$$
Under this assumption it follows that inside each subspace ${\cal H}_{n,n-2h}$
there exists a complete set of eigenfunctions of $L_m$, that its eigenvalues
$\lambda$ are real, and that the eigenfunctions are of the form
$$
D_{l_{2h+1}...l_n}({\cal P}_1\otimes...\otimes {\cal P}_1\otimes {\cal P}_
{n-2h,S})^{l_1l_2...l_n}_{i_1i_2...i_n}|i_1i_2...i_n>\in {\cal H}_{n,n-2h},
                                                              \eqno (6.41)
$$
where the coefficients $D$'s satisfy condition (6.37) with $k=n-2h$. Summing
up, for any set of coefficients $\{m_{ij}\}$ satisfying condition
(6.40) $L_m$ is a well-defined observable on ${\cal H}$
commuting with the hamiltonian and the square angular momentum:
$$
[H_{\omega},L_m]=0,~~~~~~~~~[L^2,L_m]=0.                      \eqno (6.42)
$$
For $q=1$ and a suitable normalization it coincides with a particular
component of the angular momentum. When $N=3$ $H_{\omega},L^2,L_m$ make up
a complete set of commuting observables, at least in a neighbourhood of
$q=1$, since this is the case when $q=1$. In a forthcoming paper [18] we will
study the spectrum and eigenfunctions of the operators $L_m$.

{}~~

{}~~

\centerline{\bf {\large 7}. Positivity of the scalar product}

{}~~

{}~~~In this section we prove the positivity of the
scalar product $(~~,~~)$;
in this way the proof that ${\cal H}$ is a pre-Hilbert space
is finished. Then completion of ${\cal H}$ can be performed in
the standard way. The section is not essential for a conceptual understanding
of the work and may be
skipped by the reader if he/she is not interested in computations.

{}~~

{\bf Proposition 7.1:} $\forall q\in {\bf R^+}$ the scalar product introduced
in section 5 is positive definite:
$$
(u,u)\ge 0~~~~~~~~~(u,u)=0~\Leftrightarrow~u=0,~~~~~~u\in {\cal H}.
                                                                \eqno (7.1)
$$
$Proof:$ The results of the preceding section imply that it is sufficient to
prove positivity inside each subspace ${\cal H}_{n,n-2m}$. The most general
$|u>\in {\cal H}_{n,k}$, $k=n-2m,~~0\le m\le {n\over 2}$ is of the form
$$
\matrix
{& & & & D_{l_1l_2...l_k}\psi_{n,(k,S)}^{l_1l_2...l_k}& &&&&&& \cr
 & & {\Pi \atop \nearrow} & & & &&&&&&                         \cr
 & u & & & &&&&&&D_{l_1...l_k}\in {\bf C}&                     \cr
 & & {\bar\Pi \atop \searrow} & & & &&&&&&                     \cr
 & & & & D_{l_1l_2...l_k}\bar\psi_{n,(k,S)}^{l_1l_2...l_k}& &&&&&&  \cr}
                                                                  \eqno (7.2)
$$
(see (6.25)) where
$$
\psi_{n,(k,S)}^{l_1l_2...l_k}:=(a^+_nCa^+_{n-1})...(a^+_{k+2}
Ca^+_{k+1})\psi_{k,S}^{l_1l_2...l_k}
$$
$$
\bar\psi_{n,(k,S)}^{l_1l_2...l_k}:=(\bar a^+_nC\bar a^+_{n-1})...
(\bar a^+_{k+2}C\bar a^+_{k+1})\bar\psi_{k,S}^{l_1l_2...l_k}
                                                             \eqno (7.3)
$$
and
$$
\psi_{k,S}^{l_1l_2...l_k}:={\cal P}^{~~~~~l_1...l_k}_{k,S~i_1...i_k}
\psi_k^{i_1...i_k}=t_k(q){\cal P}^{~~~~~l_1...l_k}_{k,S~i_1...i_k}
x^{i_1}...x^{i_k}exp_{q^2}[-{\omega q^{-k-N}xCx\over \mu}]
$$
$$
\bar\psi_{k,S}^{l_1l_2...l_k}:={\cal P}^{~~~~~l_1...l_k}_{k,S~i_1...i_k}
\bar\psi_k^{i_1...i_k}=t_k(q^{-1}){\cal P}^{~~~~~l_1...l_k}_{k,S~i_1...i_k}
x^{i_1}...x^{i_k}exp_{q^{-2}}[-{\omega q^{k+N}xCx\over \bar\mu}]. \eqno (7.4)
$$
Here $a_mCa_{m+1}:=a_m^iC_{ij}a^j_{m+1}$ and $a_m{+i},a_m^i$ are the
creation/destruction operators introduced in (5.2), (5.3).
A glance at formula $(5.36)_b$ and an easy calculation show
that the coefficients $t_k(q)$ $t_k(q^{-1})$
are positive $\forall q\in {\bf R}^+$.
In the rest of this section $a\propto b$ will mean $a=\sigma b,~~\sigma>0$.
The square norm of $u$ is given by
$$
(u,u)=D^*_{p_1...p_k}D_{l_1...l_k}[<(\bar \psi_{n,(k,S)}^{p_1...p_k})^*
\psi_{n,(k,S)}^{l_1...l_k}>+<(\psi_{n,(k,S)}^{p_1...p_k})^*\bar
\psi_{n,(k,S)}^{l_1...l_k}>].                                     \eqno (7.5)
$$
We are going to show our claim by proving that
 each one of the conjugate terms in the RHS of relation
(7.5) is positive.

{}~~

{\bf Lemma 7.2:}
$$
<(\bar \psi_{n,(k,S)}^{p_1...p_k})^*\psi_{n,(k,S)}^{l_1...l_k}>,
<(\psi_{n,(k,S)}^{p_1...p_k})^*\bar\psi_{n,(k,S)}^{l_1...l_k}>\propto
$$
$$
{\cal P}^{~~~~~l_1...l_k}_{k,S~i_1...i_k}{\cal P}^{~~~~p_1...p_k}
_{k,S~j_1...j_k}C^{h_1j_1}...C^{h_kj_k}\cdot
{S^{h_k...h_1i_1...i_k}\over {\cal S}_{2k}}\cdot \rho_k   \eqno (7.6)
$$
where
$$
\rho_k:=<exp_{q^{-2}}[-{\omega q^{k+N}xCx\over \bar\mu}](xCx)^k
exp_{q^2}[-{\omega q^{-k-N}xCx\over \mu}]>.               \eqno (7.7)
$$

$Proof~ of~ Lemma~ 7.2:$ From the definition $(7.3)_a$ and formulae (5.21),
(5.22),(5.36) it follows
$$
<(\bar \psi_{n,(k,S)}^{p_1...p_k})^*\psi_{n,(k,S)}^{l_1...l_k}>
\propto <(\bar \psi_{k,S}^{p_1...p_k})^*f_{k,S}^{l_1...l_k}>       \eqno (7.8)
$$
where
$$
f_{k,S}^{l_1...l_k}:=(a_kCa_{k+1})...(a_{n-4}Ca_{n-3})(a_{n-2}
Ca_{n-1})\psi_{n,(k,S)}^{l_1l_2...l_k}.                           \eqno (7.9)
$$
Because of formula (5.19), only the component of $f_{k,S}^{l_1...l_k}$
belonging to $\Psi_k$ contributes to the integral (7.8). Looking at formula
(B.10), we can decompose the operator $(a_{n-2}Ca_{n-1})$ in the following
way
$$
(a_{n-2}Ca_{n-1})=\alpha_{n-1,2}(a^+_{n+2}Ca^+_{n+1})+
\beta_{n-1,2}(a_nCa_{n+1})+\gamma_{n-1,2}(a^+_nCa_{n+1})+
$$
$$
+\delta_{n-1,2}(a_{n+2}Ca^+_{n+1}),                            \eqno (7.10)
$$
which is appropriate to clearly display the result of its action on $\Psi_n$:
we see that it maps $\psi_{n,(k,S)}$ into a combination of functions
$\psi'_{n+2},\psi'_n,\psi'_{n-2}$ belonging respectively
to $\Psi_{n+2},\Psi_n,\Psi_{n-2}$. Next, the operator $(a_{n-4}Ca_{n-3})$
acts on $\psi'_{n+2},\psi'_n,\psi'_{n-2}$. For each of these three functions
we choose the appropriate decomposition of $(a_{n-4}Ca_{n-3})$. Doing the same
job again and again, we end up with a combination of functions belonging to
$\Psi_{2n-k},\Psi_{2n-2-k},...\Psi_k$. It is not difficult to realize that
$$
{\cal P}_{\Psi_k}(f_{k,S}^{l_1...l_k})=\prod \limits_{h=1}^m
\beta_{n-2h+1,2}(a_{k+2}Ca_{k+3})...(a_n
Ca_{n+1})\psi_{n,(k,S)}^{l_1l_2...l_k},                           \eqno (7.11)
$$
where ${\cal P}_{\Psi_k}$ denotes the projector on $\Psi_k$.
Since all coefficients $\beta_{l,m}$ are
positive for $q\in {\bf R}^+$, by picking the explicit
definition $(7.3)_a$ of $\psi_{n,(k,S)}$ we find
$$
{\cal P}_{\Psi_k}(f_{k,S}^{l_1...l_k})\propto (a_{k+2}Ca_{k+3})...(a_n
Ca_{n+1})(a^+_nCa^+_{n-1})\psi_{n-2,(k,S)}^{l_1l_2...l_k}.
                                                                \eqno (7.12)
$$
In the appendix B it is proved that
$$
(a_nCa_{n+1})(a^+_nCa^+_{n-1})\psi_{n-2,(k,S)}^{l_1l_2...l_k}
\propto \psi_{n-2,(k,S)}^{l_1l_2...l_k}                         \eqno (7.13)
$$
(see formula (C.12), (C.23)); hence
$$
{\cal P}_{\Psi_k}(f_{k,S}^{l_1...l_k})\propto (a_{k+2}Ca_{k+3})...(a_{n-2}
Ca_{n-1})\psi_{n-2,(k,S)}^{l_1l_2...l_k}                      \eqno (7.14)
$$
using $m={n-k\over 2}$ times the same kind of argument we conclude that
$$
{\cal P}_{\Psi_k}(f_{k,S}^{l_1...l_k})\propto \psi_{k,S}^{l_1l_2...l_k}.
                                                               \eqno (7.15)
$$

{}~~~From eq.'s (7.8), (5.19), (7.15), (7.4), (4.15) it follows that
$$
<(\bar \psi_{n,(k,S)}^{p_1...p_k})^*\psi_{n,(k,S)}^{l_1...l_k}>\propto
<(\bar \psi_{k,S}^{p_1...p_k})^*\psi_{k,S}^{l_1...l_k}>\propto
$$
$$
\propto {\cal P}^{~~~~~l_1...l_k}_{k,S~i_1...i_k}
{\cal P}^{~~~~p_1...p_k}_{k,S~j_1...j_k}
C^{h_1j_1}...C^{h_kj_k}\cdot
$$
$$
<x^{h_k}...x^{h_1}x^{i_1}...x^{i_k}exp_{q^2}[-{\omega q^{-k-N}xCx\over \mu}]
exp_{q^{-2}}[-{\omega q^{k+N}xCx\over \bar\mu}]>=RHS(7.6)
$$
Similarly one can show the claim for
$<(\psi_{n,(k,S)}^{p_1...p_k})^*\bar\psi_{n,(k,S)}^{l_1...l_k}>$.
$\diamondsuit$

{}~

{\bf Lemma 7.3:}
$$
D^*_{p_1p_2...p_k}{\cal P}^{~~~~p_1p_2...p_k}_{k,S~j_1j_2...j_k}
D_{l_1l_2...l_k}{\cal P}^{~~~~l_1l_2...l_k}_{k,S~i_1i_2...i_k}
C^{h_1j_1}C^{h_2j_2}...C^{h_kj_k}S^{h_k...h_1i_1...i_k}~~>0.      \eqno (7.16)
$$
$Proof ~of~ Lemma~ 7.3:$ First, using property (E.6),(E.5)
of the symmetric projectors we can rewrite LHS(7.16) in the following way:
$$
[(\otimes^k C)\cdot D]^T_{j_1...j_k}
{\cal P}^{~~~~j_k...j_1}_{k,S~h_k...h_1}
{\cal P}^{~~~~l_1...l_k}_{k,S~i_1...i_k}S^{h_k...h_1i_1...i_k}
D_{l_1l_2...l_k}.
                                                               \eqno (7.17)
$$
In appendix E we prove the following relation:
$$
{\cal P}^{~~~~j_k...j_1}_{k,S~h_k...h_1}
{\cal P}^{~~~~l_1l_2...l_k}_{k,S~i_1i_2...i_k}S^{h_k...h_1i_1...i_k}=
$$
$$
=\sigma_k [{\cal P}_{k,S}\cdot (\otimes^k C){\cal P}_{k,S}]
^{l_1...l_k}_{j_1...j_k},~~~~~~~~~~~\sigma_k >0;        \eqno (7.18)
$$
a similar result could be proved in the barred case. Using property (E.6)
(E.5) again, we can rewrite the RHS of (7.17) as a sum of positive terms
$$
\sum \limits_{m_1,m_2,...,m_k} |[(\otimes^k C)\cdot{\cal P}_{k,S}\cdot
D]^{m_1m_2...m_k}|^2;                                           \eqno (7.19)
$$
this expression is always $\ge 0$ and is zero if and only if
${\cal P}_{k,S}\cdot D =0~\Leftrightarrow~u=0$ (in fact $(\otimes^k C)$
is a nondegenerate matrix) $\diamondsuit$.

{}~~

{}~~~Now we can complete the proof of Proposition 7.1: since $\rho_k>0$
(see formula (B.24)), ${\cal S}_{2k}>0$, from the two preceding lemmata
we immediately derive the thesis $\diamondsuit$.

{}~~

{}~~~Now we can introduce a norm $\Vert~~\Vert$ in ${\cal H}$ by setting
$$
\Vert u\Vert^2=(u,u).                                            \eqno (7.20)
$$
The completion $[{\cal H}]$ of ${\cal H}$ w.r.t. this norm can be performed
in the standard way. It induces completions $[V],~[\bar V]\subset
Fun({\bf R_q^N})$ of $V,\bar V$. It would be interesting to investigate if the
latter can be characterized in an intrinsic way, e.g. by characterizing their
(formal) power expansion in $x^i$'s. This is left as a possible subject for
some future work.

{}~~

{}~~

\centerline {\bf {\large 8}. Conclusions}

{}~~

{}~~

{}~~~We have shown that the quantum harmonic oscillator on ${\bf R}^N$ with
symmetry $SO(N,{\bf R})$ admits a $q$-deformation into the harmonic
oscillator on the quantum space ${\bf R}^N_q$ with symmetry $SO_q(N,{\bf R})$,
for any $q\in {\bf R}^+$.

{}~~~In fact this $q$-deformed harmonic oscillator has a lower bounded energy
spectrum; generalizing the classical algebraic construction, the Hilbert
space of physical states is built applying construction operators to the
(unique) ground state. The scalar product is strictly
positive for any $q\in {\bf R}^+$. Observables are defined as hermitean
operators, as usual. In particular we have constructed the observables
hamiltonian, square angular momentum, angular momentum components,
position operators, momentum operators. As in the classical case, the first two
and any angular momentum component commute with each other; when $N=3$ they
make up a complete set of observables.

{}~~~Both spectra of the hamiltonian and of the square angular momentum are
discrete, and the eigenvalues have the same degeneracy as in the non-deformed
case. The $q$-deformed eigenvalues are invariant under the replacement
$q\rightarrow q^{-1}$ and can be obtained from the classical ones essentially
by the replacement $n\rightarrow [n]_q$, where $[n]_q$ is the $q$-deformed
integer $n$ given by $[n]_q:={q^n-q^{-n}\over q-q^{-1}}$. Energy levels are
no more equidistant; their difference increases with $n$.

{}~~~Guiding ideas for the construction were $SO_q(N,{\bf R})$-covariance
and correspondence principle in the classical limit $q\rightarrow 1$. Essential
tools were the two differential calculi on ${\bf R}^N_q$,~ the corresponding
(and coinciding) two integrations on this quantum space and the corresponding
two representations of the Hilbert space into
the space of functions on ${\bf R}^N_q$. A sort of quantized scaling property
of the integral under dilatation of the integration variables and
correspondingly a quantization of the dimensional constant $\omega$ (the
characteristic constant of the harmonic oscillator) has been
singled out.

{}~~

{}~~

\centerline {\bf Appendix A}

{}~

{}~~~We prove formulae (4.9), (4.10) by induction.
For $n=1$ (4.9), (4.10) are true, since
$\Delta x^{i_1}x^{i_2}|=\mu \partial^{i_1}x^{i_2}=\mu C^{i_1i_2}$,
$\bar\Delta x^{i_1}x^{i_2}|=\bar\mu \bar\partial^{i_1}x^{i_2}=
\bar\mu C^{i_1i_2}$. Now assume that they are true for $n=m-1$. Then
$$
\Delta^mx^{i_1}x^{i_2}...x^{i_{2m}}|=\mu \Delta^{m-1}\partial^{i_1}
x^{i_2}...x^{i_{2m}}|+ q^2\Delta^{m-1}x^{i_1}\Delta x^{i_2}...x^{i_{2m}}|=
$$
$$
=\mu(1+q^2) \Delta^{m-1}\partial^{i_1}x^{i_2}...x^{i_{2m}}|+
q^4\Delta^{m-2}x^{i_1}\Delta^2 x^{i_2}...x^{i_{2m}}|=
$$
$$
=..................=
$$
$$
=\mu m_{q^2} \Delta^{m-1}\partial^{i_1}x^{i_2}...x^{i_{2m}}|+
q^{2m}x^{i_1}\Delta^m x^{i_2}...x^{i_{2m}}|.                 \eqno (A.1)
$$
The second term in the last expression is zero, since the $2m$ derivatives
contained in $\Delta^m$ act on $(2m-1)$ coordinates $x$; using the definition
(4.2) of the tensor $M_{2m}$, the induction hypothesis and the definition (4.5)
of $S_{2m}$ we are able to rewrite the first term as
$$
\mu m_{q^2} M_{2m,j_3...j_{2m}}^{i_1i_2...i_{2m}}\Delta^{m-1}x^{j_3...j_{2m}}|=
(\mu)^m m_{q^2}! M_{2m,j_3...j_{2m}}^{~~~i_1i_2...i_{2m}}S^{j_3...j_{2m}}
_{2(m-1)}=
$$
$$
=(\mu)^m m_{q^2}! S^{i_1i_2...i_{2m}}_{2m},                  \eqno (A.2)
$$
which shows that (4.9) is true also for $n=m$. In a similar way one proves
(4.10).

{}~~

{}~~

\centerline{\bf Appendix B}

{}~

{}~~~In this appendix we first show how to evaluate integrals of the
type
$$
<(xCx)^m exp_{q^{-2}}[-{\omega q^{N+k}xCx \over \bar\mu}]
exp_{q^2}[-{\omega q^{-N-k}xCx \over \mu}]>                      \eqno (B.1)
$$
taking $f_0:=exp_{q^{-2}}[-{\omega q^NxCx \over \bar\mu}]
exp_{q^2}[-{\omega q^{-N}xCx \over \mu}]$ as reference function. The outcoming
results, together with formulae (4.15),(4.16), will allow the determination
of all integrals involved in the scalar products of vectors of ${\cal H}$.
Second, we give some results concerning the action of creation/destruction
operators on functions $\psi\in V$.

{}~~~We start from
$$
<(xCx)^m exp_{q^{-2}}[-{\omega q^{k+N-2}xCx \over \bar\mu}]
exp_{q^2}[-{\omega q^{-k-N}xCx \over \mu}]>=
$$
$$
=q^{N+2m}<(xCx)^m exp_{q^{-2}}[-{\omega q^{k+N}xCx \over \bar\mu}]
exp_{q^2}[-{\omega q^{2-k-N}xCx \over \mu}]>                  \eqno (B.2)
$$
which is a direct consequence of the scaling property (4.23) of the integrals.
Using the q-derivative properties (2.30) of the exponentials to expand
the functions
$exp_{q^{-2}}[-{\omega q^{N-2+k}xCx \over \bar\mu}]$,

$exp_{q^2}[-{\omega q^{2-N-k}xCx \over \mu}]$ we find
$$
<(xCx)^{m+1}exp_{q^{-2}}[-{\omega q^{k+N}xCx \over \bar\mu}]
exp_{q^2}[-{\omega q^{-k-N}xCx \over \mu}]>=({q^{1-{N\over 2}}+
q^{{N\over 2}-1}\over \omega})\cdot
$$
$$
\cdot[{N\over 2}+m]_q {[m-k]_q\over [2(m-k)]_q}<(xCx)^m
exp_{q^{-2}}[-{\omega q^{k+N}xCx \over \bar\mu}]
exp_{q^2}[-{\omega q^{2-k-N}xCx \over \mu}]>,                 \eqno (B.3)
$$
i.e.
$$
<(xCx)^mexp_{q^{-2}}[-{\omega q^{k+N}xCx \over \bar\mu}]
exp_{q^2}[-{\omega q^{-k-N}xCx \over \mu}]>=
$$
$$
=({q^{1-{N\over 2}}+q^{{N\over 2}-1}\over \omega})^m
[{N\over 2}+m-1]_q[{N\over 2}+m-2]_q...[{N\over 2}]_q {[m-k]_q!\over [2(m-k)
]_q!!}\cdot
$$
$$
<exp_{q^{-2}}[-{\omega q^{k+N}xCx \over \bar\mu}]
exp_{q^2}[-{\omega q^{-k-N}xCx \over \mu}]>.                       \eqno (B.4)
$$

Now consider the integral $<f_0>$. If $k=2l$, upon use of
of the q-derivative properties (2.30) one finds
$$
<f_0>=<exp_{q^{-2}}[-{\omega q^{2l+N}xCx \over \bar\mu}]
exp_{q^2}[-{\omega q^{-2l-N}xCx \over \mu}]\cdot
$$
$$
\cdot[\prod \limits_{h=0}^{l-1}
(1-q^{2(h-l)-N}\omega{q^2-1\over \mu}xCx)][\prod \limits_{h=0}^{l-1}
(1-q^{2(l-h)+N}\omega{q^{-2}-1\over \bar\mu}xCx)]>.            \eqno (B.5)
$$
Expanding the products contained in the square brackets and using formula
$(B.4)$ to evaluate all the integrals one finds
$$
<f_0>=z_k<exp_{q^{-2}}[-{\omega q^{k+N}xCx \over \bar\mu}]
exp_{q^2}[-{\omega q^{-k-N}xCx \over \mu}]>                    \eqno (B.6)
$$
with a suitable constant $z_k$. If $k$ is even, this formula, together with
$(B.4)$, allows to evaluate any integral $(B.1)$ in terms of $<f_0>$ (which
is taken as the normalization factor of the integral). If k is odd, by
repeating the previous steps we obtain
$$
<f'_0>=z'_k<exp_{q^{-2}}[-{\omega q^{k+N}xCx \over \bar\mu}]
exp_{q^2}[-{\omega q^{-k-N}xCx \over \mu}]>,                    \eqno (B.7)
$$
where $f_0':=exp_{q^{-2}}[-{\omega q^{N+1}xCx \over \bar\mu}]
exp_{q^2}[-{\omega q^{-N-1}xCx \over \mu}]$.
Following the line suggested at the end of sect. 4, it is possible to find the
constant $\phi(q)$ such that
$$
<exp_{q^{-2}}[-{\omega q^{1+N}xCx\over \bar\mu}]
exp_{q^2}[-{\omega q^{-1-N}xCx\over \mu}]>=
$$
and
$$
\phi(q)<exp_{q^{-2}}[-{\omega q^NxCx\over \bar\mu}]
exp_{q^2}[-{\omega q^{-N}xCx\over \mu}]>                  \eqno (B.8)
$$
and to show that it is positive $\forall q\in {\bf R}^+$. We don't perform
here this computation, but just notice that by continuity the positivity of
$\phi$ must hold at least in a neighbourhood of $q=1$, since $\phi(1)=1$.
In this way all the integrals $(B.1)$ are evaluated in terms of the
normalization
constant $<f_0>$. Note that from the definition $(B.8)$ it follows
$\phi(q^{-1})
=\phi(q)$.

\centerline {**********}

{}~~~From the definition (5.2), (5.3) of the creation/destruction operators
it immediately follows
$$
{a^i_n\over d_n(q)}={q^{n+N}+q^{2-n-m}\over q^{n+N+m}+q^{2-n-m}}{a^i_{n+m}\over
d_{n+m}(q)}+{q^{n+N+m}-q^{n+N}\over q^{n+N+m}+q^{2-n-m}}{a^{+i}_{n+m}\over
b_{n+m}(q)}~~~~~~m\in{\bf Z},                                 \eqno (B.9)
$$
whence
$$
a_{n-1}Ca_n:=a_{n-1}^iC_{ij}a^j_n=\alpha_{n,m}(q)(a^+_{n+m+1}Ca^+_
{n+m})+\beta_{n,m}(q)(a_{n+m-1}Ca_{n+m})+
$$
$$
+\gamma_{n,m}(q)(a^+_{n+m-1}C
a_{n+m})+\delta_{n,m}(q)(a_{n+m+1}Ca^+_{n+m}),                \eqno (B.10)
$$
where $\beta_{n,m}(q)$ is positive $\forall q\in {\bf R^+}$.

{}~~

\centerline {**********}

{}~~~We know that
$$
(a_nCa_{n+1})(a^+_nCa^+_{n-1})\psi_{n-2,(k,S)}^{l_1...l_k}=
v_{n-2,k}\psi_{n-2,(k,S)}^{l_1...l_k}                       \eqno (B.11)
$$
(the function $\psi_{n-2,(k,S)}~~(k=n-2m)$ was defined in (7.3)), since both
sides are eigentuntions of $h_{\omega},G_{q^4}{\cal L}^2$
with the same eigenvalues and have the same transformation properties under
the coaction of the quantum group $SO_q(N,{\bf R})$. We now show that the
constant $v_{n-2,k}$ is positive $\forall q\in {\bf R^+}$. In the sequel by
$a\propto b$ we will mean $a=\sigma b$ with $\sigma>0$.
Note that $\psi_{n-2,(k,S)}^{l_1...l_k}$ can be written in the form
$$
\psi_{n-2,(k,S)}^{l_1...l_k}=[c(xCx)^{m-1}+...]exp_{q^2}[-{q^{-n-N+2}\omega xCx
\over \mu}]{\cal P}^{~~~l_1...l_k}_{k,S~i_1...i_k}x^{i_1}...x^{i_k},
                                                               \eqno (B.12)
$$
where (as in the sequel)
the dots in the square bracket denote lower degree powers of $(xCx)$. The
strategy will be to find out $v_{n-2,k}$ by only looking at the term of highest
degree in $xCx$ at each step of the derivation. From the definition (5.2),(5.3)
of the creation/destruction operators and the definition (4.18)
of the $B$ operator we get
$$
(a_nCa_{n+1})(a^+_nCa^+_{n-1})=q^{-3}[xCx+{q^{2(n+1+N)}\over
\omega^2}\Delta+{q^{n+2N}\mu^2\over \omega(q^2-1)}B+const.]\cdot
$$
$$
[xCx+{q^{10-2n}
\over \omega^2}\Delta-{q^{2-n+N}\mu^2\over \omega(q^2-1)}B+const.]G_q^4
d_n(q)d_{n+1}(q)b_n(q)b_{n-1}(q)                                 \eqno(B.13)
$$
The $\Delta$'s in the first and second square bracket have to act respectively
on functions belonging to $G_{q^2}\Psi_n$ and to $G_{q^4}\Psi_{n-2}$, therefore
they can be respectively replaced by $(q^{-N-4}\omega^2xCx-q^{-N-2}E_n)$ and
$(q^{-N-8}\omega^2xCx-q^{-N-4}E_{n-2})$. Hence
$$
(a_nCa_{n+1})(a^+_nCa^+_{n-1})\propto E\cdot F,                 \eqno (B.14)
$$
where
$$
E:=[xCx(1+q^{N+2(n-1)})+{q^{n+2N}\mu^2\over \omega(q^2-1)}B+const.]
$$
$$
F:=[xCx(1+q^{2(1-n)-N})-{q^{2-n+N}\mu^2\over \omega(q^2-1)}B+const.]G_q^4.
                                                                   \eqno (B.15)
$$
{}From formulae (6.19) and $(6.21)_a$ one easily derives the identity
$$
B{\cal P}^{~~~l_1...l_k}_{k,S~i_1...i_k}x^{i_1}...x^{i_k}|=
{q^{2k}+q^{2-N}\over \mu}{\cal P}^{~~~l_1...l_k}_{k,S~i_1...i_k}x^{i_1}...
x^{i_k}                                                           \eqno (B.16)
$$
Using the fundamental property (4.19) of $B$, formulae $(B.12)$ and $(B.16)$
we find
$$
F\psi_{n-2,(k,S)}^{l_1...l_k}=[c(xCx)^mq^{4-4m}(1+q^{2(1-n)-N})
exp_{q^2}[-{q^{-n-N-2}\omega xCx\over \mu}]+
$$
$$
-c{\mu q^{4-n-2m+N}(q^{2k}+q^{2-N})\over \omega (q^2-1)}(xCx)^{m-1}
exp_{q^2}[-{q^{-n-N}\omega xCx\over \mu}]+...]\cdot
$$
$$
{\cal P}^{~~~l_1...l_k}_{k,S~i_1...i_k}x^{i_1}...x^{i_k};         \eqno (B.17)
$$
applying the q-derivative property (2.30) to the exponential
$exp_{q^2}[-{q^{-n-N}\omega xCx\over \mu}]$ we find
$$
F\psi_{n-2,(k,S)}^{l_1...l_k}\propto [c(xCx)^m+...]exp_{q^2}[-{q^{-n-N-2}\omega
xCx
\over \mu}]{\cal P}^{~~~l_1...l_k}_{k,S~i_1...i_k}x^{i_1}...x^{i_k}.
                                                                 \eqno (B.18)
$$
After similar steps one can see that the result of the action of
$E$ on $\psi_{n-2,(k,S)}$ is
$$
E\cdot F\psi_{n-2,(k,S)}^{l_1...l_k}\propto e[c(xCx)^{m+1}+...]exp_{q^2}
[-{q^{-n-N-2}\omega xCx\over \mu}]
{\cal P}^{~~~l_1...l_k}_{k,S~i_1...i_k}x^{i_1}...x^{i_k}         \eqno (B.19)
$$
where
$$
e:=(1-q^{2m})(1-q^{2(n-m-1)+N}).
                                                                 \eqno (B.20)
$$
Using again the q-derivative property (2.30), we increase by 4 the degree of
the $q$-power in the exponent and we lower by 2 the degree of the polynomial
in $(xCx)$ contained in the square bracket, with the result that eq. $(B.11)$
holds with $v_{n-2,k}$ given by
$$
v_{n-2,k}\propto e({-\mu \over \omega(q^2-1)q^{-n-N-2}})
({-\mu \over \omega(q^2-1)q^{-n-N}}).                             \eqno (B.21)
$$
We see that $v_{n-2,k}>0~~~\forall q\in {\bf R}^+$.

\centerline {***********}

{}~~~Using the explicit definition of the creation/destruction operators and
the Schroedinger equation for $\psi_n\in \Psi_n$ it is straightforward to
show that
$$
a^+_nCa_{n+1}\psi_n=\sigma
BG_{q^2}\psi_n~~~~~~~~a_{n+2}Ca^+_{n+1}\psi_n=\sigma'
BG_{q^2}\psi_n,~~~~~~~~\sigma,\sigma'>0;                        \eqno (B.22)
$$
if $\psi_n$ is the scalar eigenfunction of level $n=2m$, then $BG_{q^2}$ acts
as the identity operator and therefore
$$
a^+_nCa_{n+1}\psi_n=\sigma\psi_n~~~~~~~~a_{n+2}Ca^+_{n+1}\psi_n=\sigma'
\psi_n,~~~~~~~~\sigma,\sigma'>0.                        \eqno (B.23)
$$

\centerline {***********}

{}~~~We show that
$$
<exp_{q^{-2}}[-{\omega q^{k+N}xCx\over \bar\mu}](xCx)^k
exp_{q^2}[-{\omega q^{-k-N}xCx\over \mu}]>~~>0~~~~~~~~~~~\forall q\in {\bf R}^+
                                                                  \eqno (B.24)
$$
First we consider the case $k=2h$. Using the scaling property (4.21) of the
integral we find
$$
LHS(B.24)=q^{-k(k+1)}<\bar \psi_0(xCxG_{q^2})^k \psi_0>.           \eqno (B.25)
$$
Looking back at formulae $(5.33)_a$, (5.42)
it is easy to prove that $(xCx)G_{q^2}$ can be decomposed in the following way
$$
(xCx)G_{q^2}={q+q^{-1}\over \omega(1+q^{-2-N})}[(a_nCa_{n+1})+(a_n^+Ca_{n+1})+
(a_{n+2}Ca^+_{n+1})+(a^+_{n+2}Ca^+_{n+1})]                  \forall n\ge 0.
                                                                  \eqno (B.26)
$$
Only the component ${\cal P}_{\Psi_0}((xCxG_{q^2})^k \psi_0)$ belonging to
$\Psi_0$ of the function $(xCxG_{q^2})^k \psi_0$
gives a nonvanishing contribution to the integral (B.25), because of property
(5.19). Using the decomposition (B.26) with $n=0,2,...,2(k-1)$, properties
(B.11),(B.21),(B.23) we see that
$$
{\cal P}_{\Psi_0}((xCxG_{q^2})^k \psi_0)=\tau_k \psi_0,             \eqno
(B.27)
$$
where $\tau_k$ is given by a sum of positive constants ($\forall q\in
{\bf R}^+$). This proves (B.24) in the case $k=2l$.

{}~~~If $k=2l+1$ an analogous reduction shows that
$$
<exp_{q^{-2}}[-{\omega q^{k+N}xCx\over \bar\mu}](xCx)^k
exp_{q^2}[-{\omega q^{-k-N}xCx\over \mu}]>=
$$
$$
=\tau'_k<exp_{q^{-2}}[-{\omega q^{1+N}xCx\over \bar\mu}](xCx)
exp_{q^2}[-{\omega q^{-1-N}xCx\over \mu}]>,                  \eqno (B.28)
$$
where $\tau'_k>0~~\forall q\in {\bf R}^+$. formulae $(B.3),~(B.8)$ imply
$$
<exp_{q^{-2}}[-{\omega q^{1+N}xCx\over \bar\mu}](xCx)exp_{q^2}[-{\omega
q^{-1-N}xCx\over \mu}]>=
$$
$$
={q^{2-N\over 2}+q^{N-2\over 2}\over \omega}[{N\over2}]_q {1\over
[2]_q}\phi(q)<exp_{q^{-2}}[-{\omega q^NxCx\over \bar\mu}]
exp_{q^2}[-{\omega q^{-N}xCx\over \mu}]>.                 \eqno (B.29)
$$
Since $\phi(q)$ is positive $\forall q\in {\bf R}^+$, $(B.24)$ is
proved for any $k$.

{}~~

{}~

\centerline {\bf Appendix C}

{}~

{}~~~In this appendix we show that any scalar polynomial $I(x,\partial)$
(resp. $I(x,\bar\partial)$) in $x^i,\partial^j$
(resp. $x^i,\bar\partial^j$) can be expressed as a polynomial in the (ordered)
variables $xCx,\Delta$ (resp. $xCx,\bar\Delta$) alone. We limit ourselves to
the unbarred case; the proof for the barred case is a word by word repetition
of the proof of the former, after obvious replacements.

{}~~~To be a scalar $I$ must be a polynomial in scalar variables of
the type
$$
\tilde I_{2n}(\varepsilon_i,\varepsilon'_j)=
(\eta_{\varepsilon_1})^{i_1}(\eta_{\varepsilon_2})^{i_2}...(\eta_
{\varepsilon_n})^{i_n}(\eta_{\varepsilon'_n})_{i_n}...(\eta_{\varepsilon'_2})
_{i_2}\eta_{\varepsilon'_1,i_1},                         \eqno (C.1)
$$
where $\varepsilon_i,\varepsilon'_j=+,-$, $\eta_+:=x$ and $\eta_-:=\partial$.
{}From here we see that $I$ can only contain terms of even degree in $\eta^i_
\varepsilon$; we denote by $I_{2m}$ a scalar polynomial of degree $2m$ and
containing only even powers of $\eta^i_{\varepsilon}$.
The only four independent  $I_2$ are $1,xCx,\Delta,x^i\partial_i$, and they
all can be expressed as polynomials in $xCx,\Delta$ because of formula (2.32).

{}~~~Our claim amounts to showing that for any $I_{2m}$ ($m\ge 0$) there exist
an
ordered polynomial $P_I(xCx,\Delta)$ in $xCx,\Delta$ such that
$$
I_{2m}=P_I(xCx,\Delta)                                  \eqno (C.2)
$$
The claim is obviously true for $m=0$. The general proof is by induction:
assume that it is true for $m=k$. Since any $I_{2(k+1)}$ can be written as
a polynomial in $\tilde I_{2n}$ variables with $n\le k+1$, it is sufficient
to prove the claim for a $\tilde I_{2(k+1)}$ whatsoever. By the induction
hypothesis and the very definition (C.1) of the $\tilde I$ variables
$\tilde I_{2(k+1)}$  can be written in the form
$$
\tilde I_{2(k+1)}=(\eta_{\varepsilon})^i \tilde P(xCx,\Delta)
(\eta_{\varepsilon'})_i                                           \eqno (C.3)
$$
with some polynomial $\tilde P$. Decomposing the latter in a sum of
monomials and using formulae
$$
\partial^i(xCx)=\mu x^i+q^2(xCx)\partial^i~~~~~~~~~
x^i\Delta=q^{-2}\Delta x^i -\mu q^{-2}\partial^i                 \eqno (C.4)
$$
to move the $\eta^i$'s
step by step through all the factors $xCx,\Delta$ as far as the extreme
right we will be able to write the RHS of (C.3) as a combination of terms
of the type $\tilde P'(xCx,\Delta)\cdot (\eta_{\varepsilon''})^i
(\eta_{\varepsilon'})_{i}$; but $(\eta_{\varepsilon''})^i(\eta_{\varepsilon'})
_i$
is a polynomial of the type $I_2$ for which the claim (C.2) holds, hence
it holds also for $\tilde I_{2(k+1)}$ and the statement (C.2) is completely
proved.

{}~~

\centerline {\bf Appendix D}

	~~

{}~~~In this appendix we list a few properties of the
projectors ${\cal P}_k^S$ defined in formula (6.8).

In a forthcoming paper we will show that
$$
{\cal P}_{k,S}=\pi_k(\hat R_{i,i+1},{\cal P}^1_{i,i+1}),~~~~~~i=1,2,...,k-1
                                                                 \eqno (D.1)
$$
$$
{\cal P}_{k,S}^T={\cal P}_{k,S}                                  \eqno (D.2)
$$

where $\pi_k$ is a polynomial in $\hat R_{i,i+1},{\cal P}^1_{i,i+1}$
such that
$$
\pi_k(\hat R_{k-i,k-i+1},{\cal P}^1_{k-i,k-i+1})=
\pi_k(\hat R_{i,i+1},{\cal P}^1_{i,i+1}).                        \eqno (D.3)
$$
Let us denote by $P_k$ the permutator on $\otimes ^k {\bf C^N}$ defined by
$$
P_k(v_1\otimes v_2\otimes ... \otimes v_k)=v_k\otimes...\otimes v_2\otimes v_1,
{}~~~~~~~~~~~~v_i\in {\bf C^N};                                     \eqno (D.4)
$$
Using the relation (2.8) it is easy to check that
$$
P_k\cdot(\otimes^k C)\hat R_{i,i+1}=\hat R_{k-i,k+1-i}P_k\cdot(\otimes^k C)
{}~~~~~~~~~~P_k\cdot(\otimes^k C){\cal P}^1_{i,i+1}=
{\cal P}^1_{k-i,k+1-i}P_k\cdot(\otimes^k C)                       \eqno (D.5)
$$
Relations (D.3), (D.5) imply
$$
[P_k\cdot(\otimes^k C),{\cal P}_{k,S}]=0                         \eqno (D.6)
$$
{}~~~Here we give, as an example, the explicit form of ${\cal P}_{3,S}$ in
terms
of $\hat R_{i,i+1},({\cal P}_1)_{i,i+1}$:
$$
{\cal P}_{3,S}={1\over 3_{q^2}!}\{ {\bf 1}+q(\hat R_{12}+\hat R_{23})+
q^2(\hat R_{12} \hat R_{23}+
\hat R_{23}\hat R_{12}) + q^3\hat R_{12}\hat R_{23}\hat R_{12}+
$$
$$
-{(q^N-1)\mu \over 2(q^{N+2}-1)}[(2q^2+1-q^4)(({\cal P}_1)_{12}+({\cal P}_1)
_{23})+2q^4Q_N(({\cal P}_1)_{12}({\cal P}_1)_{23}+({\cal P}_1)_{23}({\cal P}_1)
_{12})
$$
$$
q^2(q+q^{-1})(\hat R_{12}({\cal P}_1)_{23} +({\cal P}_1)_{23}\hat R_{12}+
\hat R_{23}({\cal P}_1)_{12}+({\cal P}_1)_{12}\hat R_{23})+
$$
$$
+(\hat R_{12}({\cal P}_1)_{23}\hat R_{12}+\hat R_{23}({\cal P}_1)_{12}
\hat R_{23})]\}.                                                 \eqno (D.7)
$$

{}~~

\centerline {\bf Appendix E}

{}~

{}~~~We give a brief proof of relation (7.18). From relation (4.9) we infer
that
$$
{\cal P}^{~~~~j_k...j_1}_{k,S~h_k...h_1}
{\cal P}^{~~~~l_1l_2...l_k}_{k,S~i_1i_2...i_k}S^{h_k...h_1i_1...i_k}=
$$
$$
=\sigma'_k {\cal P}^{~~~~j_k...j_1}_{k,S~h_k...h_1}{\cal P}^{~~~~l_1...l_k}
_{k,S~i_1...i_k}\Delta^k
x^{h_k}...x^{h_1}x^{i_1}...x^{i_k}|,~~~~~~~\sigma'_k>0.
                                                                  \eqno (E.1)
$$
Using relations (6.18),(6.19),(6.20) we can rewrite the RHS in the
following way:
$$
\sigma'_k {\cal P}^{~~~~j_k...j_1}_{k,S~h_k...h_1}{\cal P}^{~~~~l_1...l_k}
_{k,S~i_1...i_k}\Delta^{k-1}(\mu\partial^{h_k}+q^2x^{h_k}\Delta)
x^{h_{k-1}}...x^{h_1}x^{i_1}...x^{i_k}|=
$$
$$
=\sigma'_k {\cal P}^{~~~~j_k...j_1}_{k,S~h_k...h_1}{\cal P}^{~~~~l_1...l_k}
_{k,S~i_1...i_k}\Delta^{k-1}[\mu k_{q^2}x^{h_k}...x^{h_2}\partial^{h_1}
x^{i_1}...x^{i_k}|+
$$
$$
q^{2k}x^{h_k}...x^{h_1}\Delta x^{i_1}...x^{i_k}|].      \eqno (E.2)
$$
The second term in the square brackets will yield a vanishing contribution.
In fact, the operator $\Delta^{k-1}$ can transform at most $(k-1)$ of the $k$
$x^{h_i}$ into $\partial^{h_i}$, and the remaining
$x^{h_i}$'s can be moved to the
left of all derivatives using property (6.18); such an expression is zero,
since
it contains a number $l>k$ of derivatives acting on $x^{i_1}...x^{i_k}$ (i.e.
on their left). Using
$k$ times the same kind of argument we end up with
$$
{\cal P}^{~~~~j_k...j_1}_{k,S~h_k...h_1}
{\cal P}^{~~~~l_1l_2...l_k}_{k,S~i_1i_2...i_k}S^{h_k...h_1i_1...i_k}=
$$
$$
\sigma''_k {\cal P}^{~~~~j_k...j_1}_{k,S~h_k...h_1}{\cal P}^{~~~~l_1...l_k}
_{k,S~i_1...i_k}[\partial^{h_k}...\partial^{h_2}\partial^{h_1}
x^{i_1}...x^{i_k}|,~~~~~~\sigma''_k>0.                      \eqno (E.3)
$$
Now let us perform the remaining derivations in the RHS of (E.3). Using
relation (2.5) it becomes
$$
\sigma''_k {\cal P}^{~~~~j_k...j_1}_{k,S~h_k...h_1}{\cal P}^{~~~~l_1...l_k}
_{k,S~i_1...i_k}[\partial^{h_k}...\partial^{h_2}
(C^{h_1i_1}x^{i_2}...x^{i_k}+q C^{h_1q}\hat R^{i_1i_2}_{qp}x^px^{i_3}...
x^{i_k}+...)|,                                                    \eqno (E.4)
$$
and using relations (6.9) it can be written in the form
$$
\sigma''_k {\cal P}^{~~~~j_k...j_1}_{k,S~h_k...h_1}{\cal P}^{~~~~l_1...l_k}
_{k,S~i_1...i_k}[\partial^{h_k}...\partial^{h_2}
k_{q^2}C^{h_1i_1}x^{i_2}...x^{i_k}|=
$$
$$
=............=
$$
$$
=\sigma_k {\cal P}^{~~~~j_k...j_1}_{k,S~h_k...h_1}{\cal P}^{~~~~l_1...l_k}
_{k,S~i_1...i_k}C^{h_1i_1}C^{h_2i_2}...C^{h_ki_k}=
$$
$$
=\sigma_k [{\cal P}_{k,S}(\otimes^k C){\cal P}_{k,S}]^{j_k...j_1}_{l_1...l_k},
{}~~~~~~~\sigma_k>0                                               \eqno (E.5)
$$
where for the last equality we have used properties (D.2),(D.6). Relation
(7.18)
is thus proved.

{}~~

{}~~

\centerline{\bf Notes}

{}~~

$^{(1)}$ To do this job one has to manage q-series. We hope to report useful
results in this direction elesewhere [17].

{}~~

{}~~

\centerline {\bf Acknowledgements}

{}~

{}~~~I would like to thank Prof. J. Wess for introducing the problem to me and
for
enlightening discussions. I also thank prof. L. Bonora for stimulating
discussions and for encouraging suggestions. I am grateful to H. O'Campo,
M. Schlieker, W. Weich and P. Nurowski for many useful and detailed
discussions.
Finally I would like to thank Prof. M. Abud for his continuous and trustful
encouragement.

{}~

{}~

\centerline {\bf References}

{}~

[1] J. Wess, Talk given on occasion of the Third Centenary Celebrations of the
Mathematische Gesellschaft Hamburg, March 1990; S. L. Woronowicz, Publ. Rims.
Kyoto Univ. {\bf 23} (1987) 117.

{}~~

[2] V. G. Drinfeld, " Quantum Groups ", Proceedings of the International
Congress of Mathematicians 1986, Vol. 1, 798; M. Jimbo, Lett. Math. Phys.
{\bf 10} (1986), 63.

{}~

[3] L. D. Faddeev, N. Y. Reshetikhin and L. A. Takhtajan, " Quantization of Lie
Groups and Lie Algebras ", Algebra and Analysis, {\bf 1} (1989) 178, translated
from the Russian in Leningrad Math. J. {\bf 1} (1990), 193.

{}~

[4] G. Mack and V. Schomerus, " Quasi-Hopf Quantum Symmetry in Quantum Theory
",
DESY 91-037 (May 1991).

{}~

[5] S. L. Woronowicz, Commun. Math. Phys. {\bf 122} (1989) 125-170.

{}~

[6] Yu. Manin, preprint Montreal University, CRM-1561 (1988); " Quantum
Groups and Non-commutative Geometry ", Proc. Int. Congr. Math., Berkeley
{\bf 1} (1986) 798; Commun. Math. Phys. {\bf 123} (1989) 163.

{}~

[7] J. Wess and B. Zumino, Nucl. Phys. Proc. Suppl. {\bf 18B} (1991) 302.

{}~

[8] W. Pusz and S. L. Woronowicz, Reports in Math. Phys. {\bf 27} (1990) 231.

{}~

[9] J. Schwenk and J. Wess, ``A q-deformed Quantum Mechanical Toy Model '',
Preprint MPI-Ph/92-8.

{}~

[10] A. Connes, Noncommutative Differential Geometry, Publ. Math. IHES {\bf 62}
(1986) 41; and references therein.

{}~

[11] U. Carow-Watamura, M. Schlieker and S. Watamura, Z. Phys. C Part. Fields
{\bf 49} (1991) 439.

{}~

[12] O. Ogievetsky, Lett. Math. Phys. {\bf 24} (1992), 245.

{}~

[13] O. Ogievetsky and B. Zumino, Lett. Math. Phys. {\bf 25} (1992), 121.

{}~

[14] G. Fiore, " $SO_q(N,{\bf R})$-Symmetric Harmonic Oscillator on the $N$-
dim Real Quantum Euclidean Space ", Sissa preprint 35/92/EP, to be published
in the Int. J. Mod. Phys.

{}~~

[15] See for example:

A. J. Macfarlane, J. Phys. A: Math. Gen. {\bf 22} (1989) 4581;
L. C. Biedenharn, J. Phys. A: Math. Gen. {\bf 22} (1989) L873;
Chang-Pu Sun and Hong-Chen Fu, J. Phys. A. Math. Gen. {\bf 22} (1989) L983.

See also: M. Arik, Z. Phys. C {\bf 51} (1991), 627-632; A. Kempf,
" Quantum Group-Symmetric Fock Spaces with Bargmann-Fock representation ",
preprint LMU-TPW 92-4.

{}~

[16] W. Weich, " The Quantum Group $SU_q(2)$: Covariant Differential Calculus
and a Quantum-Symmetric Quantum Mechanical Model ", Ph.D. Thesis (1990),
Karlsruhe University.

{}~

[17] G. Fiore, in preparation.

{}~

[18] G. Fiore, in preparation.

\vfill\eject\end